\documentclass[11pt]{article}
\usepackage{geometry}
\usepackage{setspace}
\usepackage{graphicx}
\usepackage[export]{adjustbox}

\usepackage{comment}
\usepackage{amsmath}
\usepackage{mathabx}
\usepackage{graphicx}
\usepackage{subfigure}
\usepackage{caption}
\usepackage{subcaption}
\usepackage{mwe}
\usepackage[T1]{fontenc}

\usepackage{xcolor}
\usepackage{rotating}
\usepackage{authblk}
\usepackage{float}

\usepackage[linktoc=all,colorlinks=true,citecolor=blue,linkcolor=blue]{hyperref}

\newcommand{\subf}[2]{%
  {\small\begin{tabular}[t]{@{}c@{}}
  #1\\#2
  \end{tabular}}%
}

\title{The NHIM bifurcation scenario of a particle in an asymmetric binary system of dwarf galaxies}

\author[1]{Christof Jung \thanks{ jung@icf.unam.mx }}
\author[1,2,3]{Francisco Gonzalez Montoya \thanks{f.gonzalez.montoya@ciencias.unam.mx}}

\small{
\affil[1]{Instituto de Ciencias Físicas, Universidad Nacional Autónoma de México, Av. Universidad s/n, Col. Chamilpa,
Cuernavaca, Morelos, CP 62210, México }
\affil[2]{Facultad de Ciencias,  Universidad Nacional Autónoma de México, Av. Universidad 3000, Circuito Exterior s/n,
Coyoacán, CP 04510, Ciudad Universitaria, Ciudad de México, México}
\affil[3]{Faculty of Engineering and Physical Sciences, University of Leeds, Leeds, LS2 9JT, United Kingdom}
}

\begin{document}

\maketitle

\begin{abstract}

We study the bifurcation scenario of a three-degree-of-freedom Hamiltonian system, a model based on the Lagrange restricted
3-body problem: a test particle moving in the gravitational field of a pair of interacting dwarf galaxies. 
The phase space of this system has 3 fundamental normally hyperbolic invariant manifolds (NHIMs) 
and their invariant stable and unstable manifolds form homoclinic/heteroclinic tangles. 
As the perturbation parameter increases, the NHIMs begin to lose normal hyperbolicity and their constituent KAM tori break, 
creating transient chaotic dynamics around them. We also observe a certain kind of coordination between the bifurcation scenarios of these NHIMs. 
We analyse this phenomenon using Poincaré maps and the delay time function. 

\end{abstract}

\newpage

\section{Introduction}

In chaotic dynamical systems, geometrical structures in the phase space, such as unstable invariant subsets, determine
the global dynamics and, in particular, the creation of chaos to a large extent. As the simplest example, we can think of the
unstable fixed points in a 2-dimensional map
whose stable and unstable manifolds build up the chaotic invariant set, which can be displayed pictorially as a horseshoe construction
traced out by the intersecting stable and unstable manifolds of the most important unstable fixed points. In maps with a domain of
higher dimension, the role played in 2-dimensional maps by unstable fixed points is taken over by normally hyperbolic invariant manifolds
(standard abbreviation NHIMs) of codimension 2; for general information on NHIMs see \cite{wi94}. These invariant subsets have stable and
unstable manifolds of codimension 1, which can form
walls, channels, and tubes in the phase space, thereby directing the global dynamics. This is the direct generalisation of the
role of stable and unstable manifolds of hyperbolic fixed points in the phase space of 2-dimensional maps.  It is then easy to
understand that the parameter dependence of the main NHIMs greatly influences the parameter dependence of the entire dynamics.
Therefore, we are interested in a good understanding of the typical bifurcation scenarios of NHIMs under parameter changes. In
autonomous Hamiltonian systems, the most important parameter is, in many
cases, the energy $E$, i.e., the conserved numerical value of the generating function $H$ of the dynamics.

Hamiltonian systems have an effective potential; for the general recipe to construct it for a given Hamiltonian
function, see \cite{sm1, sm2, ma1}. And usually, the most important NHIMs of codimension-2 are associated with the index-1 saddle
points of this effective potential. Then, these NHIMs act as transition states of the potential energy saddles and direct the flow through
these saddles \cite{waa1, waa2, ez}. Due to the uniqueness of solutions of ODEs, in the phase space, two different trajectories cannot cross. Then
segments of the stable/unstable manifolds form cylinders that direct the trajectories through the saddle effective potential.
This property demonstrates the importance of the NHIMs for the global properties of the dynamics.

For Hamiltonian systems with three degrees of freedom (3-dof), the understanding of the typical bifurcation scenarios of codimension-2 NHIMs is far from complete. There are only mathematical theorems which state that NHIMs
are persistent under perturbations as long as the normal instability remains larger than the tangential instability.
For various versions of proofs, see \cite{fen}, \cite{bb}, chapter 3 in \cite{wi94}, and chapter 3 in \cite{eld}. 
Recent computer-assisted algorithms to prove the existence and persistence of NHIMs are described in \cite{haro} and \cite{canadell}.
Many numerical examples indicate that NHIMs break in phase space regions where the tangential instability surpasses the normal
instability. In the present situation, it is worth studying many examples of bifurcation scenarios of NHIMs numerically to gain more
information about the typical and frequent bifurcation scenarios.

In the literature, there are some investigations on the loss of normal hyperbolicity of invariant surfaces \cite{loss1,loss2,loss3,loss4}.
These publications use 2-dof systems or a simplification equivalent to the reduction to 2-dof. These works are mainly
concerned with the consequences of the loss of normal hyperbolicity to transition state theory and
to the flow rates through the saddles. However, there is not much literature about the NHIMs breaking into
(partially lower-dimensional) invariant fragments, a topic into which we will delve in the present article. For some further
observations of qualitative changes of NHIMs under parameter changes, see \cite{ch1, ch2, ch3, ch4}.

NHIMs of dimension 2 or more in the phase space have an internal dynamics which can be displayed graphically by plotting the Poincaré
map. For an early example of the use of such a Poincaré map, see Fig.1 in \cite{loss1}. It is a similar idea to the
Poincaré map for the centre manifold of the saddle point, see for example Fig.6 in \cite{jm}. For Poincaré maps with 4 dimensions in
total, the Poincaré map for NHIMs of codimension 2 is itself 2-dimensional and can be plotted conveniently by projecting it on a
canonical plane. This projected Poincaré map, as a function of the system parameters, is a simple way to display the bifurcation scenario of the NHIMs. The numerical method to follow the NHIMs under parameter changes and to display
the projected Poincaré map has been explained in detail in \cite{gj} and in the appendix.

In various examples, we have observed that in systems with several fundamental NHIMs, the bifurcation scenarios of these NHIMs are
coordinated and important qualitative changes of one particular NHIM occur in the same small parameter interval in which also some
other NHIMs experience important changes. One of these examples has been a simple model for the motion of a test particle in the
effective potential of a rotating symmetric system of two spherical dwarf galaxies \cite{zj1}. This system has three collinear
Lagrange points and codimension-2 NHIMs associated with these saddle points of the effective potential. The two outer NHIMs are symmetry-related, whereas the middle one is qualitatively different from the outer ones. Nevertheless, the bifurcation scenario of the middle
NHIM is coordinated with the bifurcation scenario of the two outer NHIMs. Now the reader can ask whether the discrete left-right symmetry of the complete system is responsible for this coordination of the bifurcation scenarios. Motivated by this question, we
investigate in the present article the asymmetric generalisation of the model from \cite{zj1}. In the present article, we set the
masses of the two spherical dwarf galaxies to different values in contrast to \cite{zj1}, where these two masses are equal.

Another example of the occurrence of coordination effects in the bifurcation scenario of NHIMs has been observed in \cite{zj2}.
This system is the motion of a test particle in a rotating double-barred galaxy. The model has four codimension-2 NHIMs associated with
index-1 saddles of the effective potential. These four NHIMs fall into two non-equivalent groups, where the two NHIMs within each
group are symmetry related. Nevertheless, we found a coordination of the scenarios of all four NHIMs, i.e. also between the
non-equivalent groups.

In the treatment of these previous systems, we did not observe any important role of the transient outer parts of partially
broken NHIMs. In the meantime, two interesting cases of such transient effects have been observed, see \cite{ju21} and
\cite{gjman}.  Interestingly, in the asymmetric case of the two dwarf galaxies, the system studied in the present article, the
transient outer parts of the broken NHIMs influence the coordination of the bifurcation scenario
and thereby of the creation and development of chaos. This is a new event that was not previously noted in previous studies.
We consider it a remarkable new result of the investigation presented here. For general information on transient dynamics and, in particular
on transient chaos, see the textbook \cite{tel}.

In section 2, we introduce the asymmetric model and explain the important extremal points of its effective potential. Section 3
displays the bifurcation scenario of the 3 NHIMs and of the most important periodic orbits related to these NHIMs. In
section 4,  we have a closer look at the creation of chaos and at heteroclinic connections between NHIMs, including the role of
transient effects in the dynamics. In section 5, we use the delay time as an indicator function to obtain more information about the phase space
structure, and in section 6, we draw conclusions and give final remarks.

\newpage

\section{The model}

In \cite{zj1}, the functional form of the model has already been given for general values of the masses $m_1$ and $m_2$ of the two
galaxies. Therefore, we give here only a brief repetition of the most important properties of the model. The total mass of the two galaxies taken
together is $m_t = m_1 + m_2$. The distance between the centres of the two galaxies is $R$. We describe the dynamics in a corotating
frame with Cartesian coordinates $x, y, z$ where the origin is the centre of mass of the whole system. The $z$ axis is the axis of rotation
with the angular velocity $\Omega$. In the rotating frame, the centres of the two individual galaxies are located along the $x$
axis at the points $x_1 = - R  m_2/m_t $ and $x_2 = R m_1/m_t $ respectively and independently of the time.
The gravitational potential for a test particle at a general point $x, y, z$ is given as
\begin{equation}
V_g = -\frac{G m_1}{ r_1} - \frac{G m_2}{r_2},
\end{equation}
where $G$ is the gravitational constant and
\begin{eqnarray}
r_1^2 = (x - x_1)^2 + y^2 + z^2 + c^2, \\
\nonumber r_2^2 = (x-x_2)^2 + y^2 + z^2 + c^2.
\end{eqnarray}
Here, $c$ is a size measure for the galaxies, which we assume to be equal for the two galaxies. The two galaxies rotate around each other
and when $R$ is large compared to $c$ then we can approximate the rotational angular velocity $\Omega$ by the Kepler value
\begin{equation} 
\Omega = \sqrt{\frac{G m_t}{ R^3}}.
\end{equation}
Accordingly, the effective potential in the rotating frame is given as
\begin{equation}
V_{eff}(x,y,z) = V_g(x,y,z) - \frac{\Omega^2}{2} (x^2 + y^2).
\end{equation}
We use the same units as in the symmetric case; they are explained in detail in section 2 of \cite{zj1}.

In these units, the parameters of the system used in the following are $c=0.25$, $R=6$, $m_1 = 0.06$, and $m_2 = 0.14$. Note that
accordingly $m_t=0.2$, exactly the same value as the one used for the symmetric case in \cite{zj1}. Also, the values of $c$
and $R$ are chosen the same as in the symmetric case. We hope that this makes a comparison between the symmetric and asymmetric
cases easier. The momenta conjugate to the Cartesian position coordinates are $p_x, p_y, p_z$. In these coordinates, the generating
function of the dynamics (the Hamiltonian function) is given as
\begin{equation}
H(x,y,z,p_x,p_y,p_z) = \frac{1}{2}(p_x^2 + p_y^2 + p_z^2 ) + V_g(x,y,z) - \Omega L_z.
\end{equation}

Here $L_z = x p_y - y p_x$ is the $z$ component of the angular momentum. We describe the dynamics in a corotating system.
Therefore, the numerical value of the function $H$ is the Jacobi constant, denoted by the letter $E_J$.

In later sections, a knowledge of some symmetries of this system will be useful. First, we notice that the effective potential is
symmetric under $y \rightarrow -y$ and under $z \rightarrow -z$. If a trajectory starts with $z=0$ and $p_z=0$, then this trajectory will
lie completely in the horizontal plane $z=0$ of the position space. Therefore, this horizontal plane is an invariant 2-dof subsystem.
Because we describe the dynamics in a rotating frame, we do not find the usual time reversal symmetry.
However, if we combine the transformation $t \rightarrow -t$ with a reflection in the plane $y=0$, then the orientation of rotation is
also reversed, and the Hamiltonian equations of motion remain the same. Therefore, the whole dynamics is invariant under the following
transformation
\begin{equation}
(t,x,y,z,p_x,p_y,p_z) \rightarrow (-t,x,-y,z,-p_x, p_y,-p_z).
\end{equation}
This transformation can be used to transform stable manifolds of invariant subsets into unstable manifolds of the corresponding $ y$-reflected
invariant subsets. This transformation will become useful in section 4.

\begin{figure}[h!]
\begin{center}
  \includegraphics[scale=1.0]{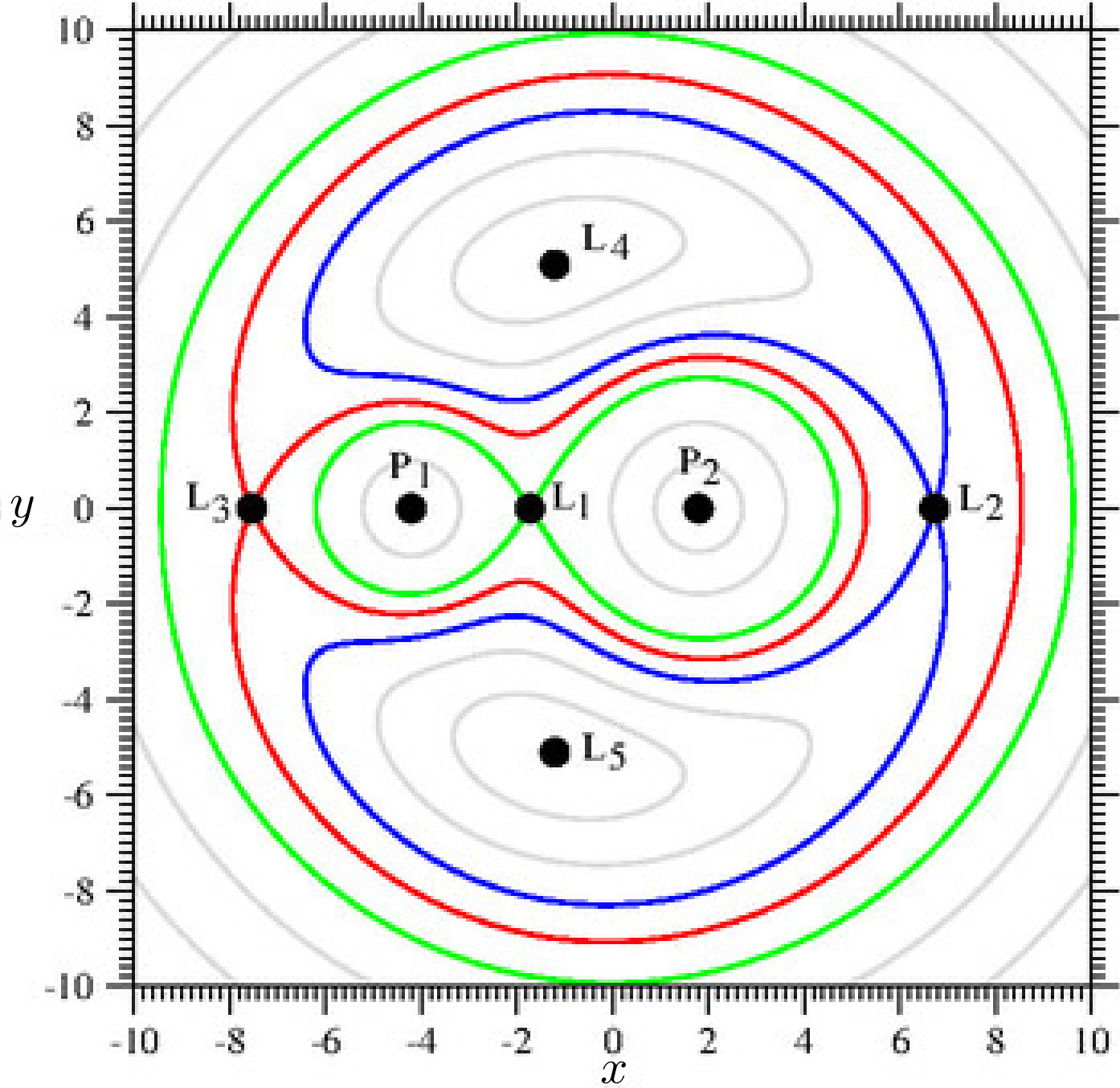}
\caption{The effective potential $V_{eff}$ in the horizontal plane. The green, blue and red curves
are the equipotential lines to the energies $E_{J1}$, $E_{J2}$, and $E_{J3}$, respectively. Some additional
equipotential curves are drawn as grey lines. The black dots mark the extremal points of the potential.   }
\label{fig:1}
\end{center}
\end{figure}

Some contour lines of the effective potential from Eq.4 are plotted in the horizontal $x-y$ plane in  Fig.\ref{fig:1}. The points $L_1$, $L_2$, and
$L_3$ are the collinear Lagrange points. In the 3-dimensional position space, they are index-1 saddles of $V_{eff}$ where
the $x$ direction is the unstable direction. The 3 important NHIMs of the system are associated with these saddle points. The Lagrange points $L_4$
and $L_5$ are index-2 saddle points of $V_{eff}$ and are of little importance in the following. The points $P_1$ and $P_2$ are relative minima
of $V_{eff}$ and they lie close to the centres of the two galaxies because the distance $R$ is large compared to the size $c$.
Compare this plot with Fig.3 in \cite{zj1}, which displays the corresponding symmetric case. In the following, we call $E_{Ji}$
the value of $V_{eff}$ in the Lagrange point $L_i$. And we use the names $E_{J6}$ and $E_{J7}$ for the values of $V_{eff}$ in the points $P_1$
and $P_2$ respectively. The numerical values of $E_{J1}$ for our choice of the masses are: 
$E_{J1}\approx-0.06511$, $E_{J2}\approx-0.05481$, $E_{J3}\approx-0.05921$,
$E_{J4}=E_{J5}\approx-0.04647$, $E_{J6}\approx-0.2714$, and $E_{J7}\approx-0.5714$.

In Fig.\ref{fig:1}, the equipotential curves to the energies $E_{J1}$, $E_{J2}$, and $E_{J3}$ are given by the green, blue, and red curves, respectively.
The equipotential curves for some further Jacobi constant values are plotted as grey curves. The extremal points of the potential are marked
by black dots and labelled by their names given in the last paragraph.

Finally, we give the motivation for our choice of the mass ratio between the two galaxies. As already mentioned, we want to keep $m_t$ fixed
at the value 0.2. And then, it is clear that we have chosen $m_1=m_2=0.1$ for the symmetric case in \cite{zj1}. And it is also obvious that in
the symmetric case we had $E_{J2}=E_{J3}$ and also $E_{J6}=E_{J7}$, see Fig.3 in \cite{zj1}. In the present article, we want to have the system strongly
asymmetric and all three $E_J$ values of the collinear Lagrange points as distinct as possible. So we have chosen the values of the masses
such that the distance between $E_{J1}$ and $E_{J3}$ is
approximately equal to the distance between $E_{J3}$ and $E_{J2}$. Now the reader might ask: Why not use $m_2$ close to 0.2 and correspondingly
$m_1=m_t - m_2$ close to zero? Then the ratio between $m_2$ and $m_1$ would be enormous. The problem is that in such a case, $E_{J3}$ and $E_{J1}$
would be very close together, and the saddle point $L_3$ would have little importance for the dynamics.

\newpage

\section{Numerical results for the bifurcation scenario of the NHIMs}

\subsection{Bifurcation scenario of the most important periodic orbits}

\begin{figure}[h!]
    \centering
    \begin{tabular}{c c}   
     \subf{\includegraphics[width=9cm]{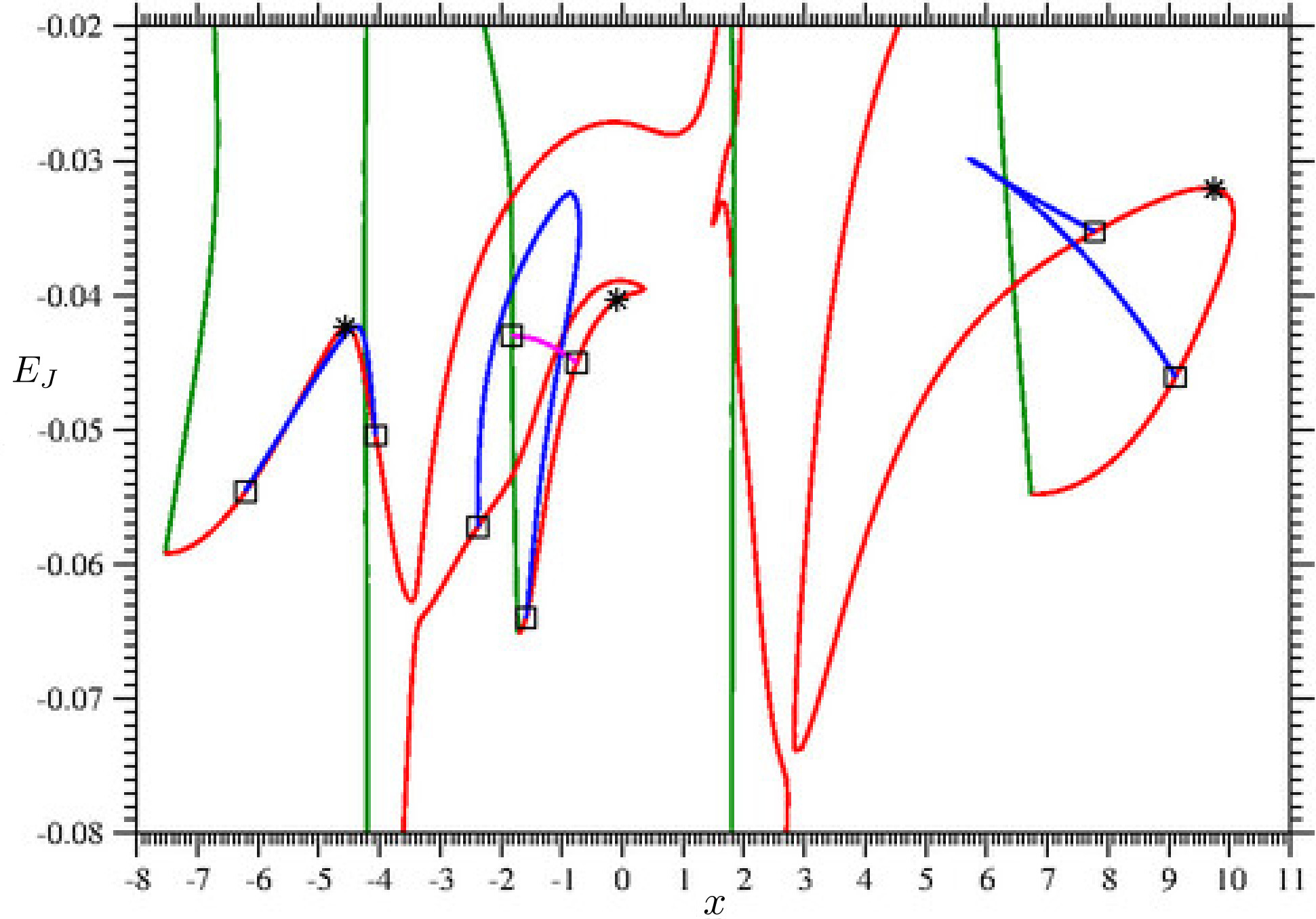}}
     {(a)}
      &
     \subf{\includegraphics[width=9cm]{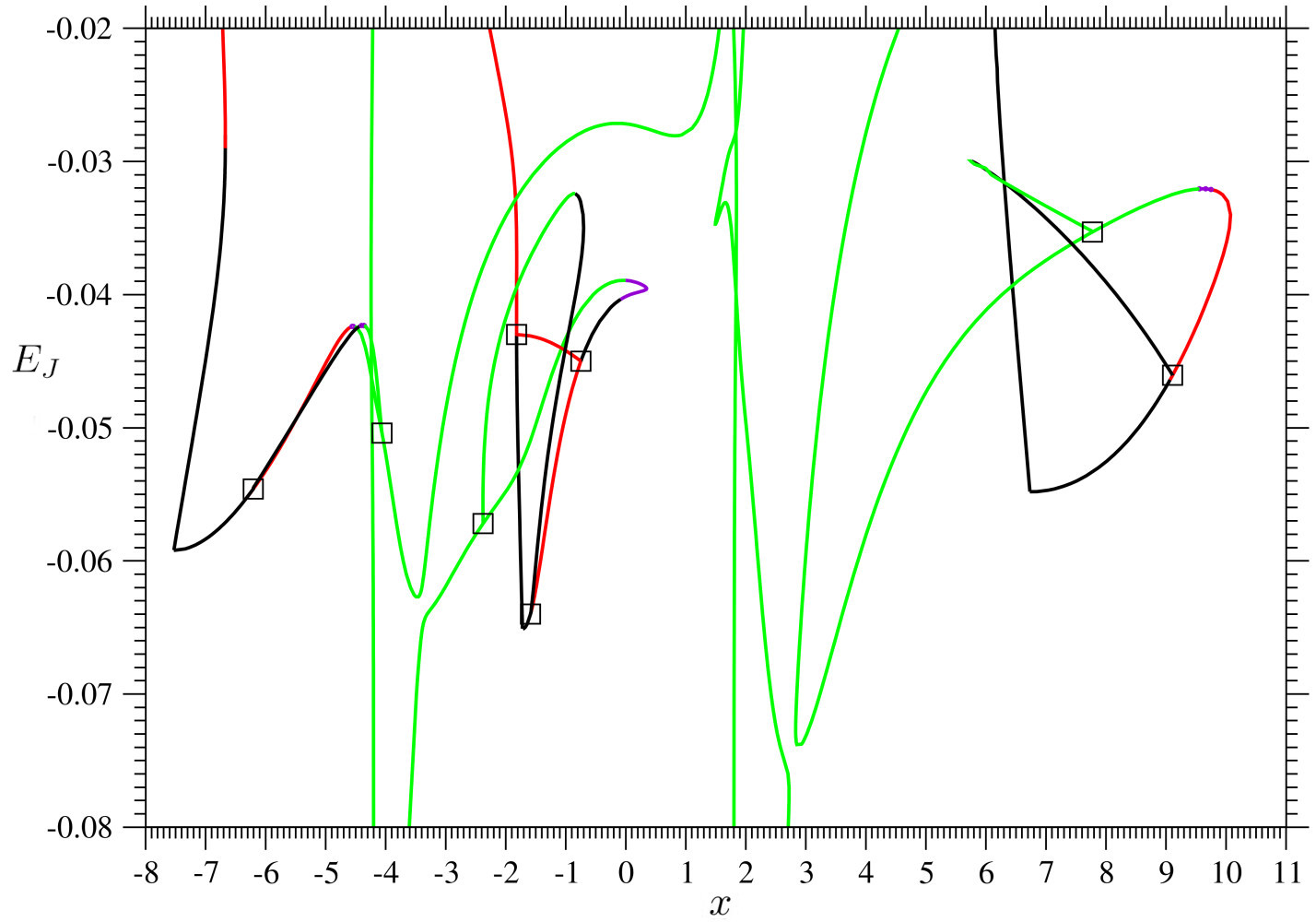}}
     {(b)}
     \\
    \end{tabular}
    \caption{ Bifurcation diagram of the most important periodic orbits. On the panel (a), red curves represent
horizontal orbits, dark green curves represent vertical orbits, blue curves represent tangentially
stable tilted loop orbits, and the magenta curve represents tangentially unstable tilted loop orbits.
Pitchfork bifurcations are marked as open black squares, and the loss of normal hyperbolicity is
marked by eight-pointed black stars. On the panel (b), black curves represent periodic orbits 
which are unstable in the normal direction and stable in the tangential
direction. Red curves represent periodic orbits which are unstable in 
the normal direction and also unstable in the tangential direction, where, however, the normal instability is larger than the tangential
instability. Violet curves represent periodic orbits with larger
tangential instability. And green curves represent periodic orbits
outside of the NHIMs.}
    \label{fig:2}
\end{figure}

From now on, we use the Jacobi constant $E_J$, the conserved numerical value of the Hamiltonian, as the perturbation parameter of the system and study the
changes in the phase space as a function of $E_J$. Out of the index-1 saddles and out of the relative minima of $V_{eff}$ grow the most important periodic orbits of the system, which are partly also the most important periodic orbits within the NHIMs.
Therefore, a detailed knowledge of the bifurcation scenario of these periodic orbits helps a lot to understand the bifurcation
scenario of the NHIMs and, in the end, the changes of the whole dynamics. The important properties of these periodic orbits are
displayed in the bifurcation diagram plotted in Fig.\ref{fig:2} in an interval of the Jacobi constant which is relevant for the topic of the
present article.

The horizontal axis of the plot gives the $x$ coordinate of the periodic orbit at the moment when it crosses the plane $y=0$
of the position space in negative orientation. The vertical axis of the plot gives the value $E_J$ of the generating function $H$.
This holds for both parts of the figure.

Out of each minimum of the potential grow three fundamental periodic orbits, two of them are horizontal, i.e. completely contained
in the plane $z=0$ of the position space, and one of them goes mainly into the vertical direction. One horizontal orbit has a
negative sense of rotation, and the other one has a positive orientation of rotation. The horizontal orbits with positive orientation are not essential to understand the change of the dynamics studied in the present work and are not discussed in the following, nor included in Fig.\ref{fig:2} in order not to overload the plot. In contrast, all the periodic horizontal orbits with a negative orientation and with period
one with respect to the negative intersection with the plane $y=0$ are central for the change of the dynamics and are included
in Fig.\ref{fig:2}(a) as the red lines. There is one red curve growing out of each potential minimum. Note that each one of these two red
curves runs through a symmetry-broken pitchfork bifurcation, whereby each one turns into 3 red curves for higher values of the
Jacobi constant. For the red curve coming out of the minimum $P_1$ this happens for $x \approx -3, E_J \approx -0.063$, and for
the one growing out of the minimum $P_2$, it happens for $ x \approx 2.8, E_J \approx -0.074$.
For the event of symmetry-breaking pitchfork bifurcations, see section 20.3 and figure 20.3.2 in \cite{wig2}.

The approximately vertical periodic orbits are represented in the Fig.\ref{fig:2}(a) as the dark green curves. Again, the $x$ value of the
orbit at the moment of the negative intersection with the plane $y=0$ is plotted as a function of $E_J$.

Out of each one of the 3 collinear Lagrange points grows one horizontal Lyapunov orbit with negative orientation of rotation
in the $x$--$y$ plane, again represented in the Fig.\ref{fig:2}(a) as red curves, and one approximately vertical Lyapunov orbit, again
represented in the Fig.\ref{fig:2}(a) as dark green curves. In the following, we call $lh_i$ and $lv_i$ the horizontal Lyapunov orbit and the
vertical Lyapunov orbit, respectively, coming out of the Lagrange point $L_i$, $i=1,2,3$. These Lyapunov orbits are the most important and
central periodic orbits within the NHIMs created from the collinear Lagrange points. And these Lyapunov orbits are created at $E_{J,i}$ as normally hyperbolic and tangentially elliptic, where the words normally and tangentially refer to the NHIM.

Pitchfork bifurcations play an important role in the bifurcation scenario, and they are indicated by open black squares in the plot.
In such a pitchfork bifurcation, each one of the horizontal Lyapunov orbits splits off a pair of tilted loop orbits, which
go out of the horizontal plane. The two members of such a pair of tilted loop orbits are transformed into each other by a
reflection in $z$ and are represented by a single blue curve in the Fig.\ref{fig:2}(a). For a higher Jacobi constant, these tilted loop
orbits disappear in saddle-centre bifurcations with pairs of further tilted loop orbits ( also represented by blue curves )
in Fig.\ref{fig:2}(a), being created in pitchfork bifurcations of the horizontal orbits coming out of the potential minima.
The tilted loop orbits of this type created from the Lagrange point $L_i$ will be called $lt_i$ in the following. The
orbit $lh_1$ runs through a further pitchfork bifurcation at $E_J \approx -0.045$, where it splits off a further pair of
tilted loop orbits which are represented by the magenta curve in Fig.\ref{fig:2}(a). These tilted loop orbits will be called $lu_1$.
With increasing Jacobi constant, these new tilted loop orbits $lu_1$ turn vertical very rapidly and are absorbed by the vertical
Lyapunov orbit $lv_1$ in an inverse pitchfork bifurcation at $E_J \approx -0.043$.

At still higher Jacobi constant, the horizontal Lyapunov orbits coming out of the collinear Lagrange points are destroyed in
saddle-centre bifurcations together with horizontal periodic orbits coming out of the potential minima or the horizontal
orbits created in the broken pitchfork bifurcations suffered by the horizontal orbits coming out of the potential minima.

The stability properties of the periodic orbits contained within the NHIMs and their changes during all these
bifurcations are essential for the bifurcation scenario of the NHIMs. Therefore, we illustrate them in Fig.\ref{fig:2}(b), which contains
the same orbits as Fig.\ref{fig:2}(a), only plotted in other colours which indicate stability properties. Black represents periodic orbits
within the NHIMs, with instability in the normal direction combined with stability in the tangential direction. Red represents
orbits within the NHIMs with instability in the normal direction and also instability in the tangential direction, where, however,
the normal instability is higher than the tangential instability. Violet indicates orbits originally growing out of the NHIMs,
where the tangential instability has already become larger than the normal instability. Accordingly, the violet orbits no longer belong to the NHIMs. Finally, green indicates orbits outside of the NHIMs, they are not part of the following analysis. 
The transition from red to violet of a curve in Fig.\ref{fig:2}(b) indicates the loss of
normal hyperbolicity of this orbit. In Fig.\ref{fig:2}(a), the loss of normal hyperbolicity
of horizontal Lyapunov orbits is marked by 8-pointed black stars. The implications of this loss of normal hyperbolicities for
the changes of the NHIMs will become evident below in the following subsections describing the bifurcation scenarios of the NHIMs.

To end this subsection, let us compare the actual Fig.\ref{fig:2} with Fig.11 in \cite{zj1} for the corresponding symmetric case.
Because of the symmetry in the old plot, the orbits coming out of $L_3$ and a large part of the orbits coming out of
$P_1$ are omitted. On the other hand, in the old plot are included a few additional orbits which have turned out to be of little interest
for the bifurcation scenario. In the new plot, those irrelevant orbits are not included in order not to overload the plot.
Therefore, to compare with the new asymmetric plot, we should imagine the old plot of the symmetric case augmented by the orbits
coming out of $L_3$ and of $P_1$ and should delete the irrelevant orbits. Then we see a strong qualitative analogy between the
two plots. There is only one essential difference which we point out now. In the new plot of Fig.\ref{fig:2} we see two rather sharp
corners in the red curves, one at $x \approx 0.3, E_J \approx -0.039$ and the other one at $x \approx 1.5, E_J \approx -0.035$.
When the two masses become more similar and approach each other, then these two corners also approach each other in the middle
and for equal masses, they touch and form a pitchfork bifurcation where $lh_1$ absorbs a pair of other horizontal orbits coming
from the potential minima. What we see in the asymmetric case in Fig.\ref{fig:2} is a symmetry-broken pitchfork instead.

\subsection{Bifurcation scenario of the internal dynamics of the NHIM $\mathcal{M}^1_{E}$}

\begin{figure}[h!]

    \centering
    \begin{tabular}{c c}   
     \subf{\includegraphics[scale=1.0 ]{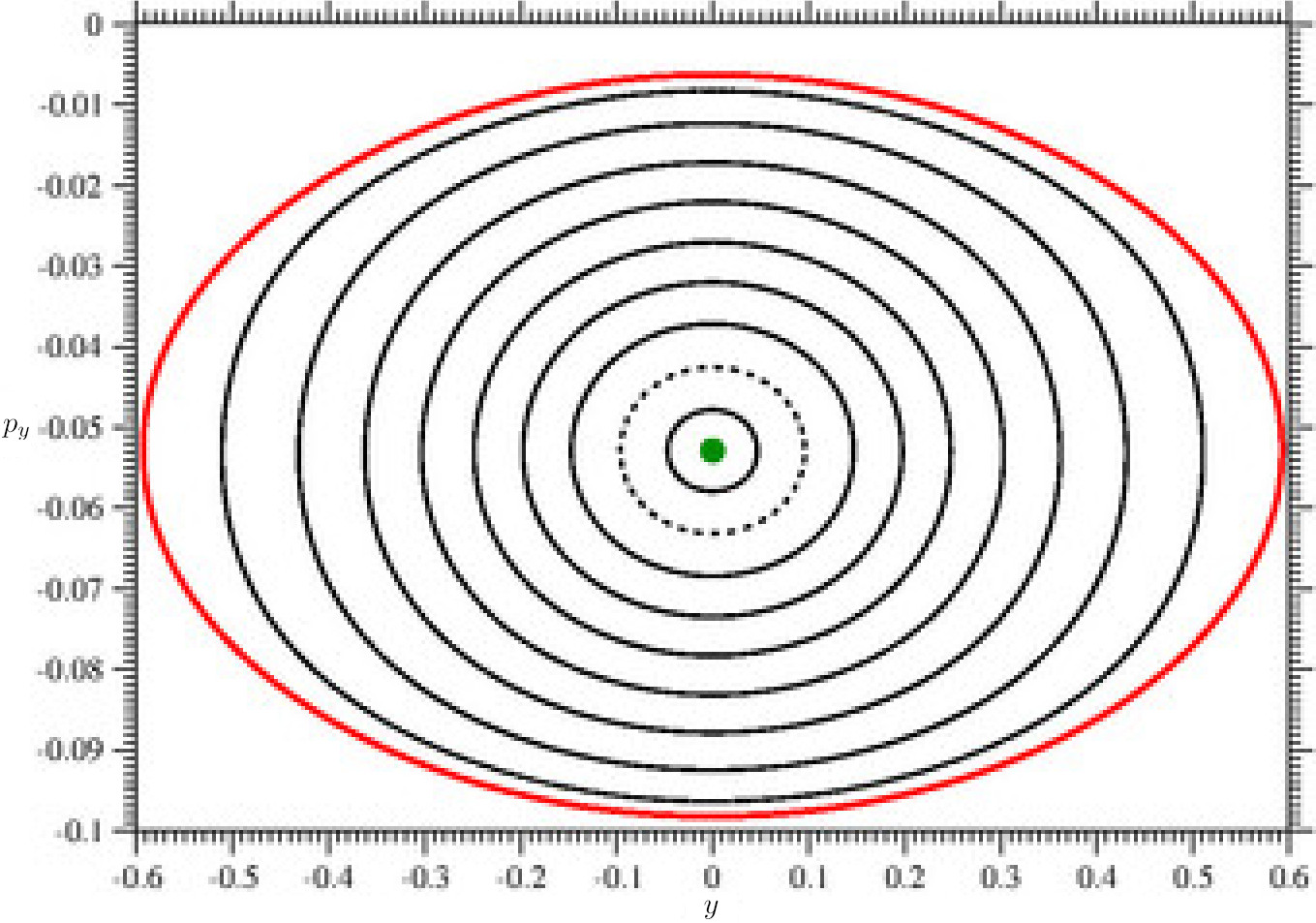}}
     {(a) $E_J = -0.064$}
      &
     \subf{\includegraphics[scale=1.0 ]{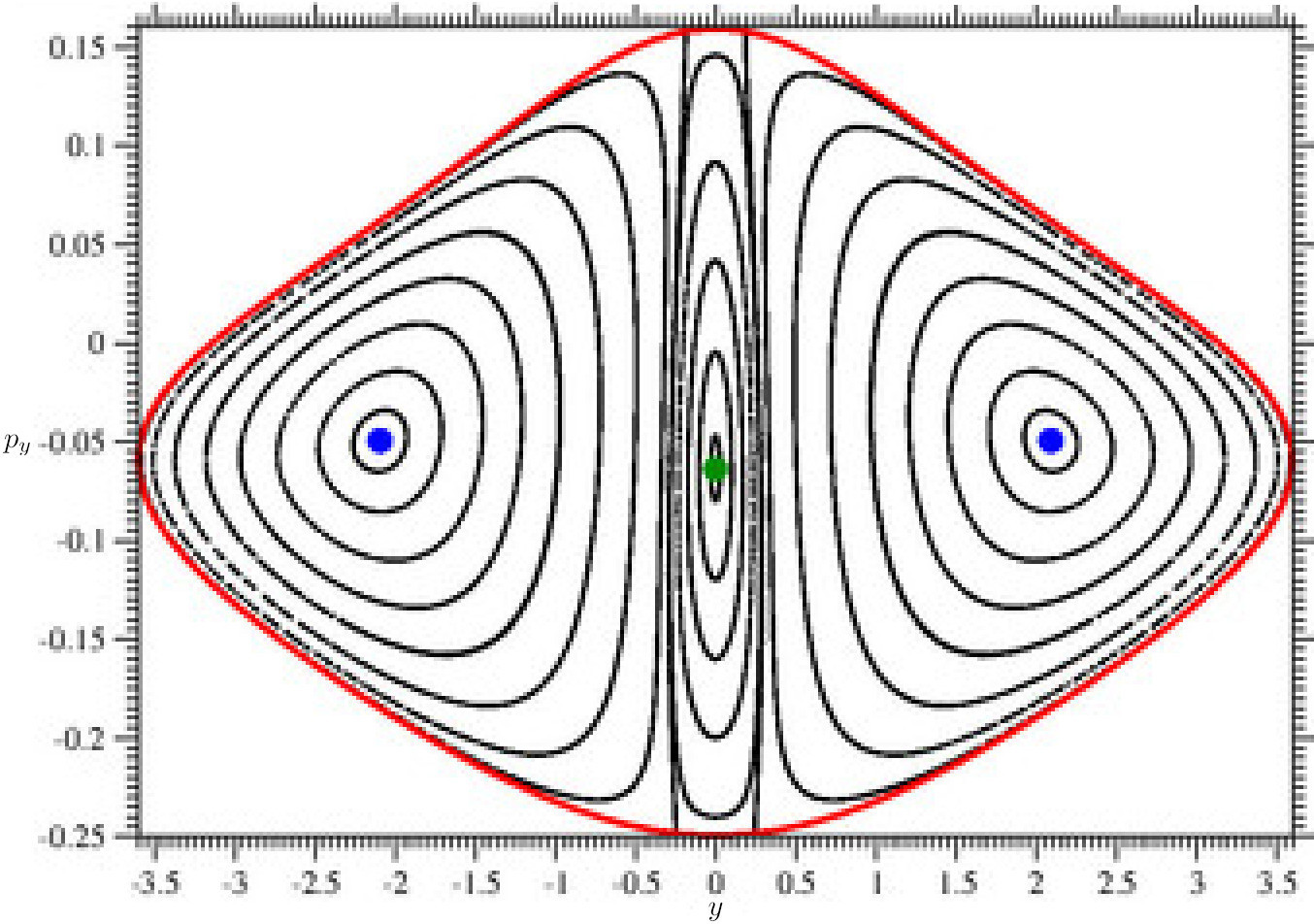}}
     {(b) $E_J = -0.046$}
     \\
     \subf{\includegraphics[scale=1.0 ]{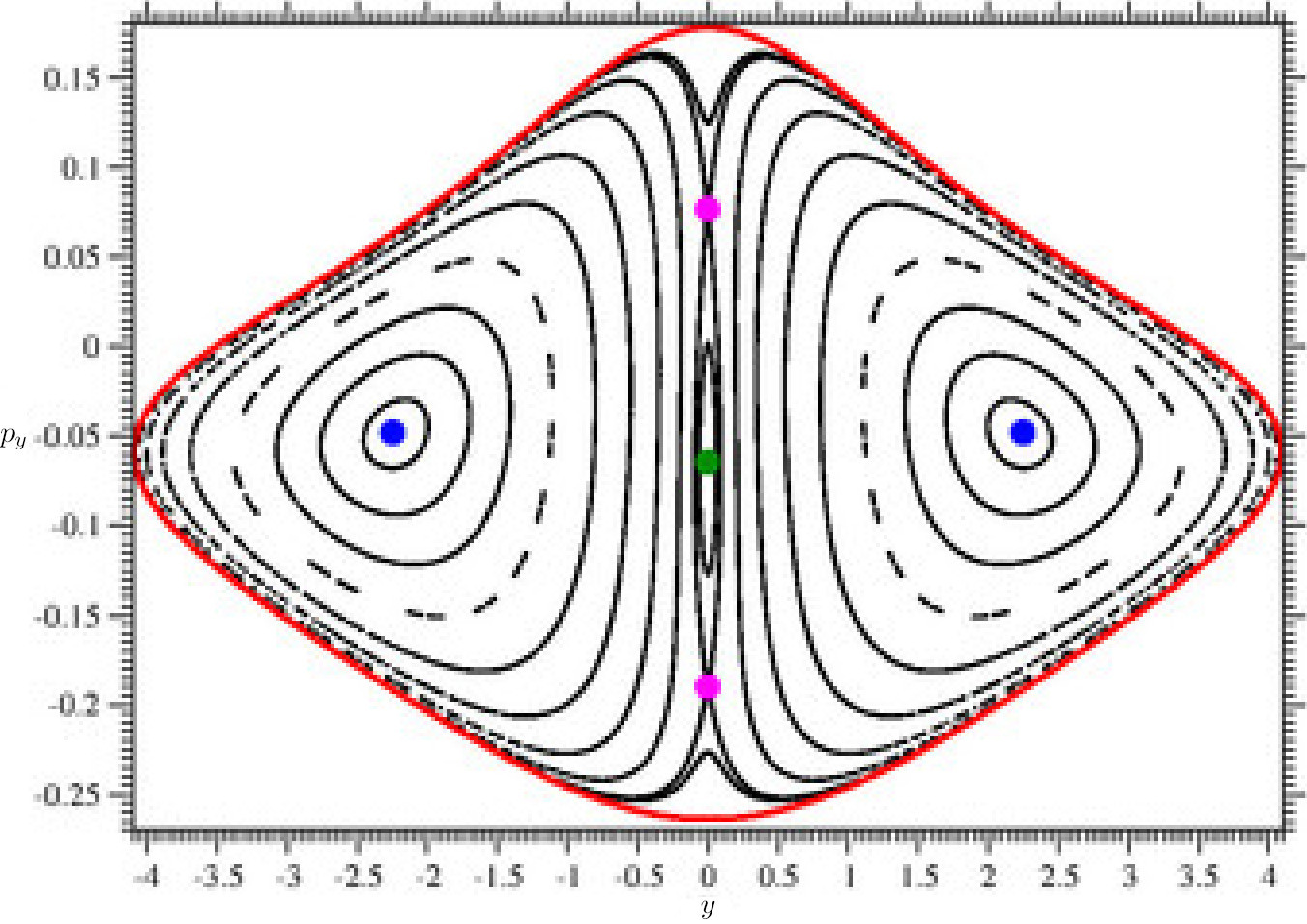}}
     {(c) $E_J = -0.044$}
     &
     \subf{\includegraphics[scale=1.0 ]{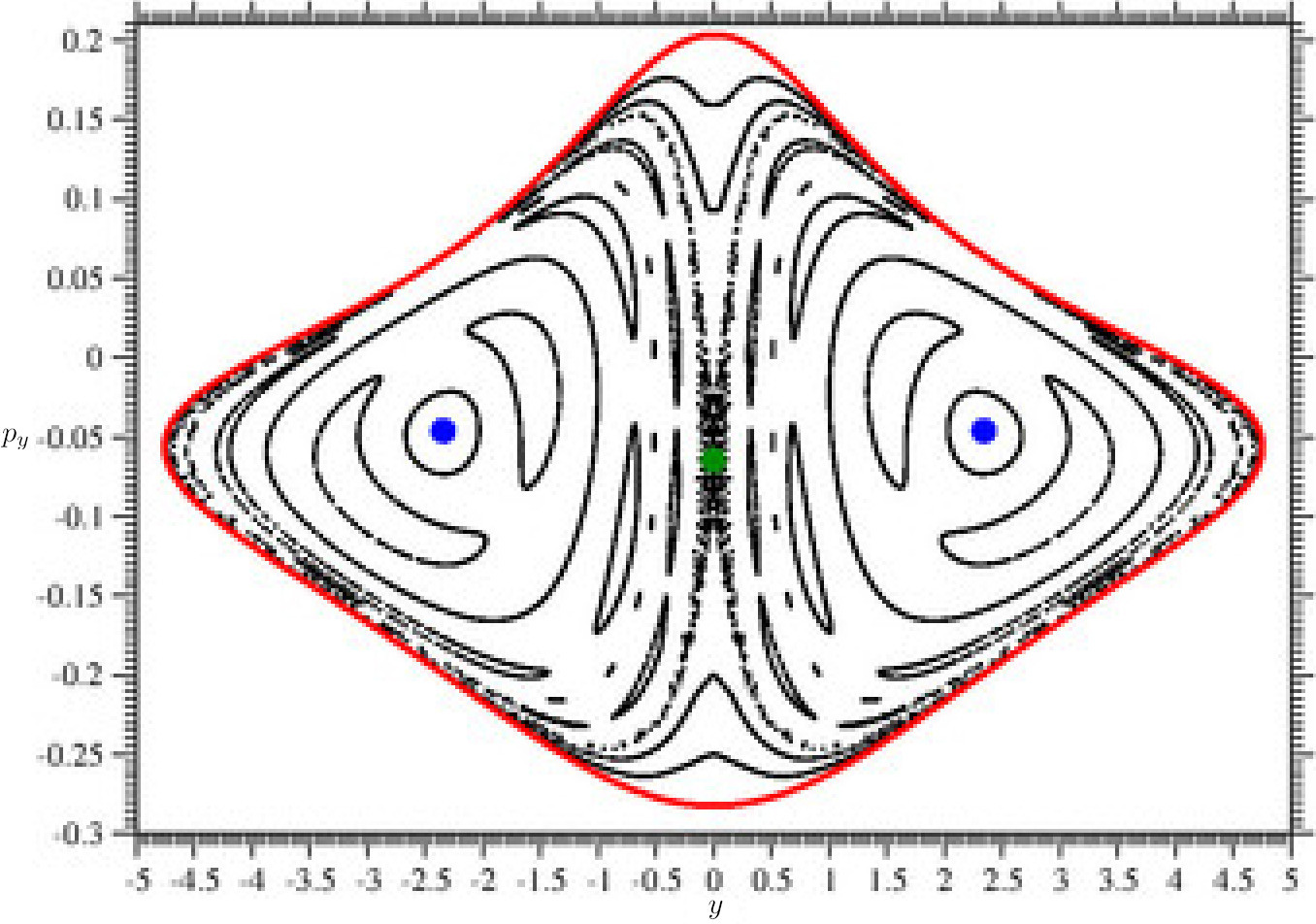}}
     {(d) $E_J = -0.042$}
     \\
     \subf{\includegraphics[scale=1.0 ]{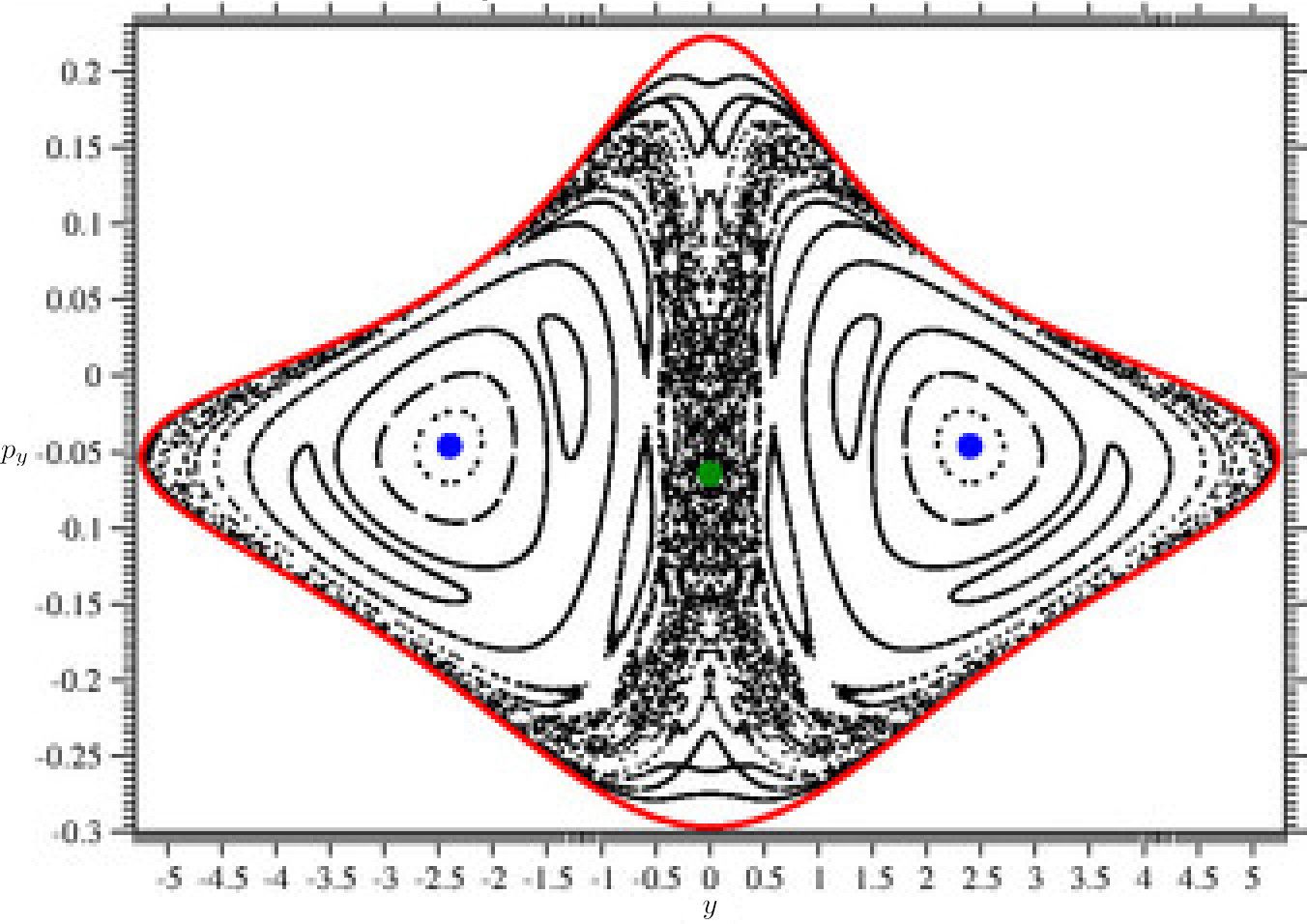}}
     {(e) $E_J = -0.041$}
     &
     \subf{\includegraphics[scale=1.0 ]{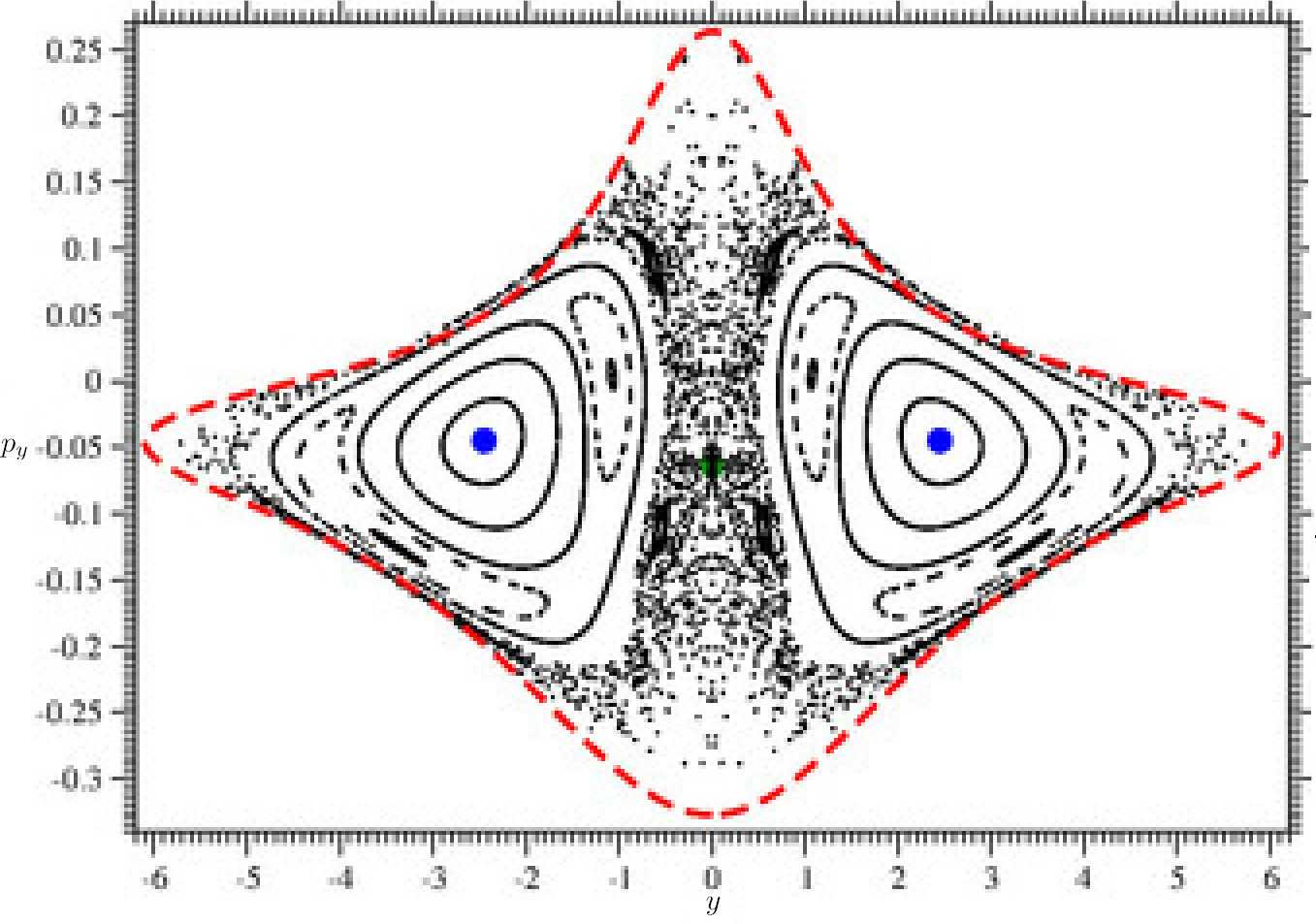}}
     {(f) $E_J = -0.04$}
    \end{tabular}
    \caption{Projected Poincaré map of the NHIM $\mathcal{M}^1_{E_J}$ and its remnant with transient trajectories for different values of the Jacobi constant $E_J$ }
    \label{fig:3}
\end{figure}

We show the internal dynamics of NHIMs using the Poincaré map of the NHIM with the intersection condition $z=0$
where the intersection orientation is $p_z<0$. In the full 4-dimensional Poincaré map,
the NHIM is a 2-dimensional curved surface embedded in the 4-dimensional domain of the
map having coordinates $x, y, p_x, p_y$. To display the Poincaré map graphically, we show the projection of the Poincaré map of the NHIM
surface into some appropriate 2-dimensional surface. Here, the choice of the $y$--$p_y$ canonical plane gave the best graphical
results. Therefore, we will use this projection for all projected Poincaré maps throughout the article.

The central and most important NHIM is the one associated with the Lagrange point $L_1$, let us denote this NHIM as $\mathcal{M}^1_{E_J}$ for the
value $E_J$ of the Jacobi constant. This NHIM
is  created at the Jacobi constant $E_{J1}$. Then, we start with the graphical presentation of the bifurcation scenario of this NHIM.
Very close to this Jacobi constant $E_{J1}$, the internal dynamics of the NHIM $\mathcal{M}^1_{E_J}$ is
a two degrees of freedom (2-dof) oscillator, and accordingly, the projected Poincaré map in Fig.\ref{fig:3}(a)  for $E_J = -0.064$ looks like the one of a
2-dof oscillator. The red solid curve displays the horizontal Lyapunov orbit $lh_1$, which lies completely within the horizontal
plane $z=0$. It marks the projection of the boundary of the NHIM $\mathcal{M}^1_{E_J}$ into the intersection plane $z=0$. When points
on the projected Poincaré map of NHIM $\mathcal{M}^1_{E_J}$ approach this red boundary curve, then the value of $p_z$ of their corresponding trajectories in the moment of intersection approaches 0.

For low energies, both of the Lyapunov orbits are tangentially elliptic. Now we increase the Jacobi constant and show plots of the
qualitatively changed NHIM $\mathcal{M}^1_{E_J}$ appearance after each important bifurcation event, which is related to an important
bifurcation event of the Lyapunov orbits, presented in the previous subsection.

At a Jacobi constant $E_J \approx -0.0638$, the periodic orbit $lh_1$ experiences its first pitchfork bifurcation, where it becomes tangentially hyperbolic and splits
off a pair of tangentially elliptic tilted loop orbits. In the projected Poincaré map, these tilted loop orbits are represented by new fixed
points which, at the bifurcation Jacobi constant, come out of the boundary, i.e. out of $lh_1$. In the figures, these tilted fixed points are
represented by blue dots. The tilted loop orbits are surrounded by KAM tori, which grow larger
with increasing Jacobi constant. Fig.\ref{fig:3}(b) shows the situation for $E_J=-0.046$. Here, the central KAM tori around $lv_1$ is already
rather small; the large majority of the NHIM $\mathcal{M}^1_{E_J}$ is occupied by the KAM tori around the tilted loop orbits. In its first pitchfork bifurcation,
$lh_1$ has become tangentially unstable and has become part of a separatrix structure which divides the various islands. Strictly speaking,
this separatrix structure is a fine chaos strip. However, it is so fine that in the figure it looks like lines. The tangential instability
of $lh_1$ is here still smaller than its normal instability. Accordingly, $lh_1$ still belongs to the NHIM $\mathcal{M}^1_{E_J}$.

At a Jacobi constant a little below $E_J = -0.045$, the periodic orbit $lh_1$ experiences its second
pitchfork bifurcation, where it splits off a new pair of tangentially hyperbolic tilted loop orbits, while $lh_1$ itself returns to tangentially
elliptic. The Fig.\ref{fig:3}(c) for the Jacobi constant $E_J=-0.044$ shows how these new tilted loop orbits run along the line $y=0$ closer to
the central point for increasing Jacobi constant. The 2 fixed points in the map, which correspond to these 2 new tilted loop orbits, are marked by
magenta dots. The unstable manifolds of the new tangentially unstable tilted loop orbits create a fine chaos strip in the map, which, in the figure, looks
like a separatrix structure. This separatrix divides all the various disjoint regular islands. First, we have the very small
remnant of the central KAM tori around $lv_1$. Then we have the island pair around the tangentially elliptic tilted loop orbits. And last,
In the projected Poincaré map, we have a new regular region of KAM curves close to the boundary (which is still $lh_1$), which has now become tangentially elliptic again.

At the Jacobi constant $E_J \approx -0.043$, the projected Poincaré map of the tangentially hyperbolic tilted loop orbits reach the central point and are absorbed by this
central point in an inverse pitchfork bifurcation. Thereby, the central point turns tangentially hyperbolic and becomes the central point of
the separatrix structure. The island around  $lv_1$ in the projected Poincaré map has been lost completely in this bifurcation.
The Fig.\ref{fig:3}(d) shows the projected Poincaré map of NHIM $\mathcal{M}^1_{E_J}$ for $E_J=-0.042$. 
We see how the separatrix structure has become a chaos strip of a size which is already clearly
visible in the plot. In addition, we notice that the tilted loop islands start to break from the outside, i.e. from the neighbourhood
of the chaos strip. Now, the system is in the Jacobi constant region, which is most important for the whole bifurcation scenario. Therefore, we show
in Fig.\ref{fig:3}(e) the projected Poincaré map of NHIM $\mathcal{M}^1_{E_J}$ for the nearby Jacobi constant $E_J=-0.041$, even though no important bifurcation event
happens during this further increase of the Jacobi constant. Now the
separatrix has already turned into a large-scale chaos region. It is important to emphasise that up to this Jacobi constant,
the NHIM $\mathcal{M}^1_{E_J}$ has not yet started its breaking process, i.e. in all points of the NHIM $\mathcal{M}^1_{E_J}$, the normal instability is
still larger than the tangential instability. This also holds for the large chaos region in the Poincaré map of NHIM $\mathcal{M}^1_{E_J}$.

At $E_J \approx -0.04035$, a very drastic event happens to the NHIM $\mathcal{M}^1_{E_J}$. Here, the Lyapunov orbit $lh_1$ loses its normal
hyperbolicity and the KAM closed invariant curves in its neighbourhood disappear. 
Accordingly, for higher values of the Jacobi constant, $lh_1$ no longer belongs to NHIM $\mathcal{M}^1_{E_J}$, and it can no
longer act as the boundary of the projected Poincaré map of NHIM $\mathcal{M}^1_{E_J}$. At this Jacobi constant, NHIM $\mathcal{M}^1_{E_J}$ starts to break.
At first sight, the plot of the projected Poincaré map of NHIM $\mathcal{M}^1_{E_J}$ for $E_J=-0.04$ in Fig.\ref{fig:3}(f) does not look very different from
the plot in Fig.\ref{fig:3}(e). Nevertheless, there are some important qualitative differences which we have to point out. As already
mentioned, $lh_1$ is no longer the boundary of the projected Poincaré map of the NHIM $\mathcal{M}^1_{E_J}$. 
Then, this orbit has been included in the plot as a broken line only. 
The large chaos strip that existed before became a region with transient chaos 
because there is no longer any KAM invariant torus which
prevents trajectories from diffusing chaotically to the outside in the long run and leaving the NHIM $\mathcal{M}^1_{E_J}$ tangentially.
Trajectories only stay temporarily in the transient chaos region around the NHIM $\mathcal{M}^1_{E_J}$. For more details on this type of transient behaviour on broken NHIMs, see \cite{ju21}.

\subsection{Bifurcation scenario of the NHIM $\mathcal{M}^3_{E_J}$}
Next, let us have a look at NHIM $\mathcal{M}^3_{E_J}$ associated with the Lagrange point $L_3$. This NHIM $\mathcal{M}^3_{E_J}$ is created at the
Jacobi constant $E_{J3} \approx -0.05921$. For energies a little
larger than $E_{J3}$, also this NHIM has an internal dynamics of a 2-dof oscillator, and accordingly, the projected Poincaré map shows the
corresponding pattern of concentric ellipses, where the central fixed point (again represented by a green dot in the figures) represents
the vertical Lyapunov orbit $lv_3$ and the boundary (again represented by a red solid curve in the figure) is the horizontal Lyapunov
orbit $lh_2$. It is not worth showing a plot for a very low Jacobi constant.

\begin{figure} [h!]
    \centering

    \begin{tabular}{c c}   
     \subf{\includegraphics[scale=1.0 ]{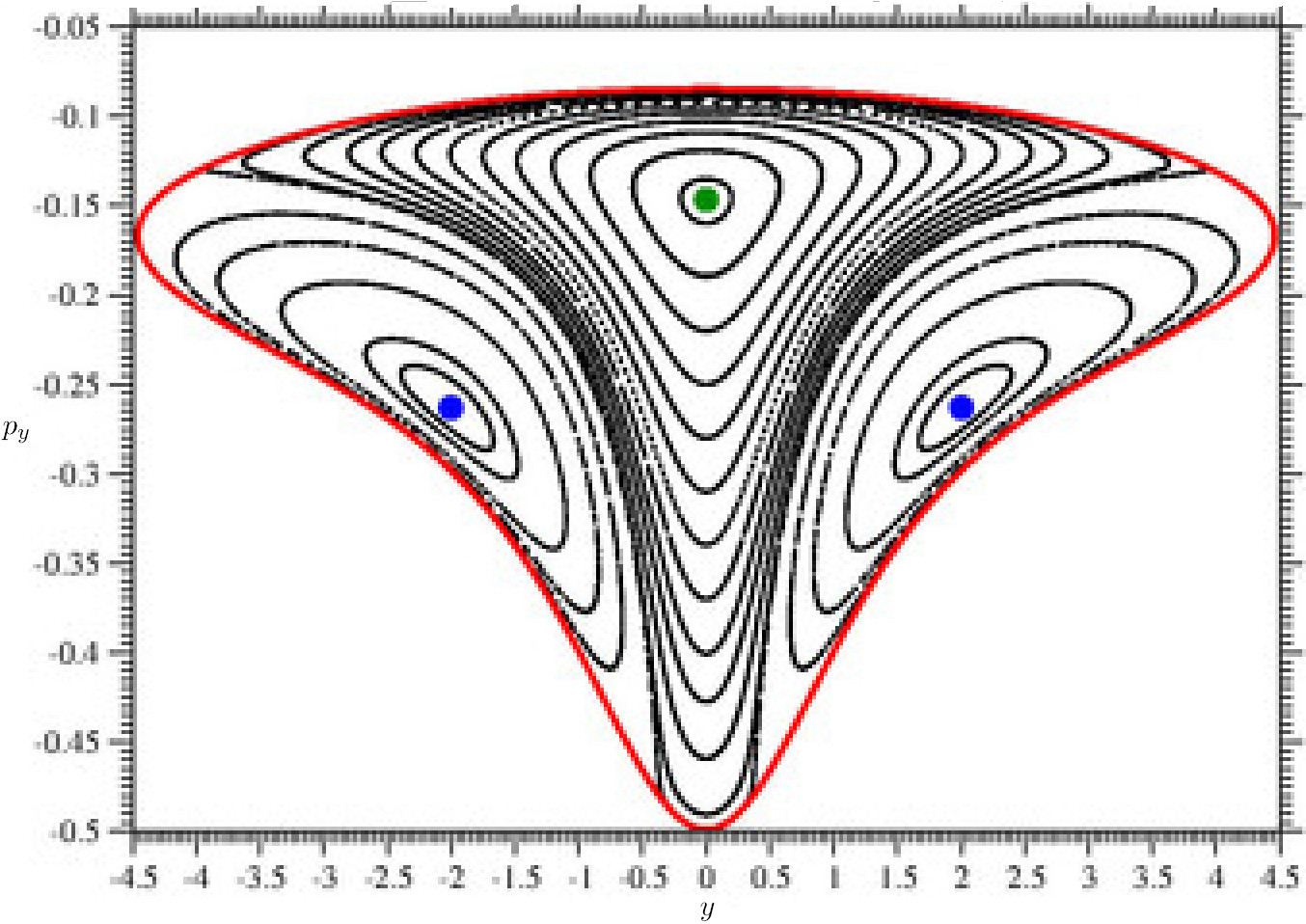}}
     {(a) $E_J = -0.045$}
      &
     \subf{\includegraphics[scale=1.0 ]{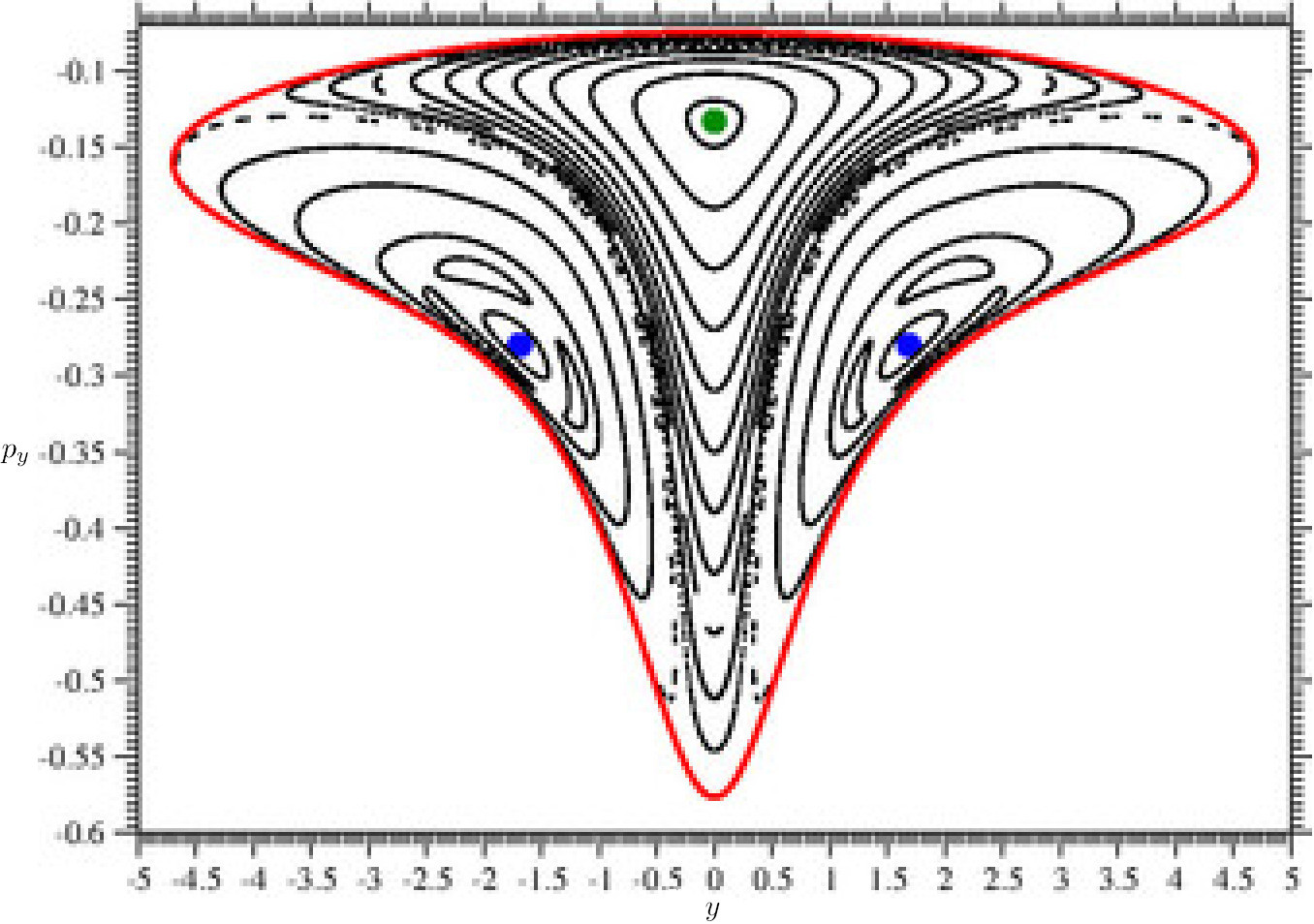}}
     {(b) $E_J = -0.043$}
     \\
     \subf{\includegraphics[scale=1.0 ]{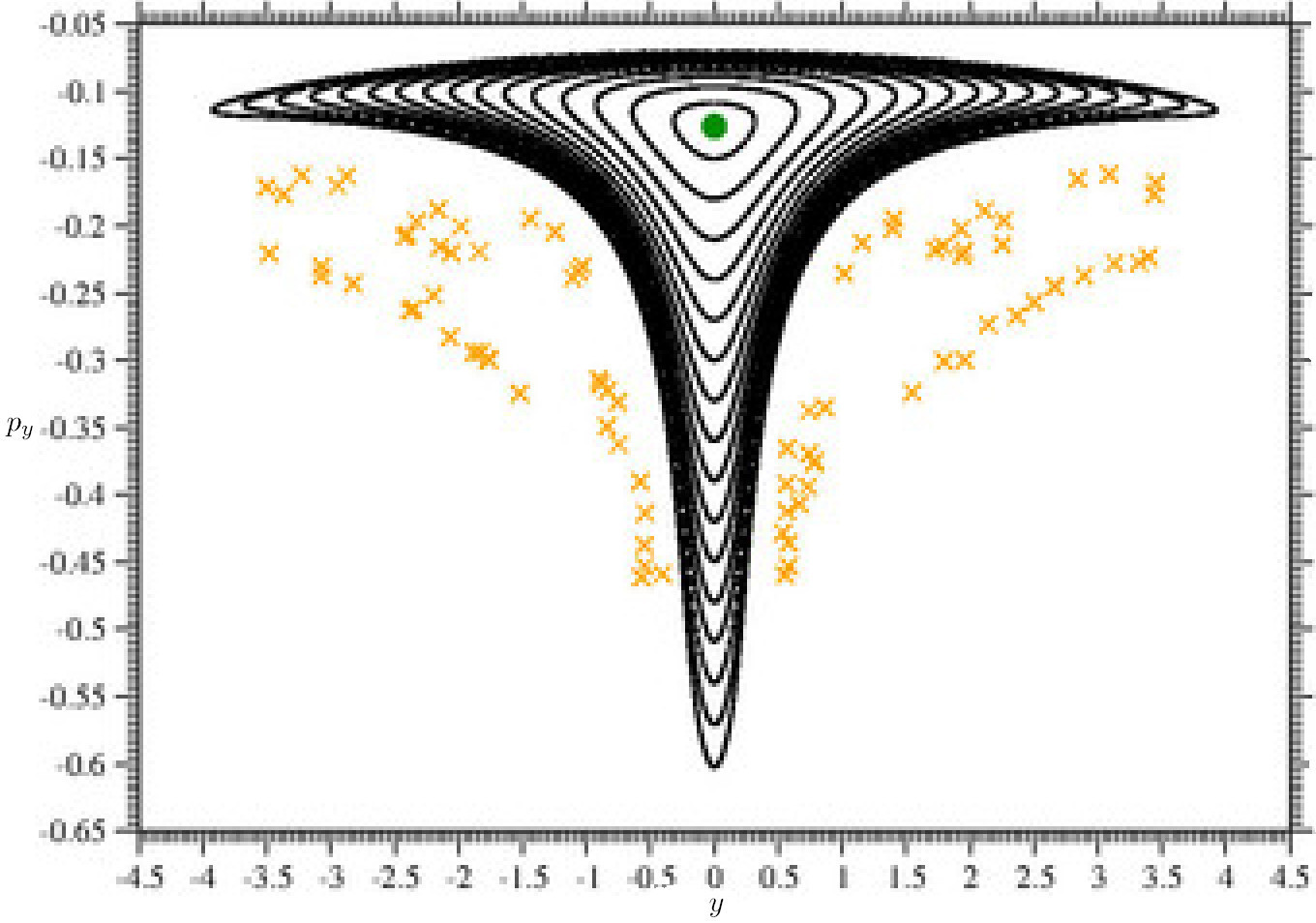}}
     {(c) $E_J = -0.042$}
     &
     \subf{\includegraphics[scale=1.0 ]{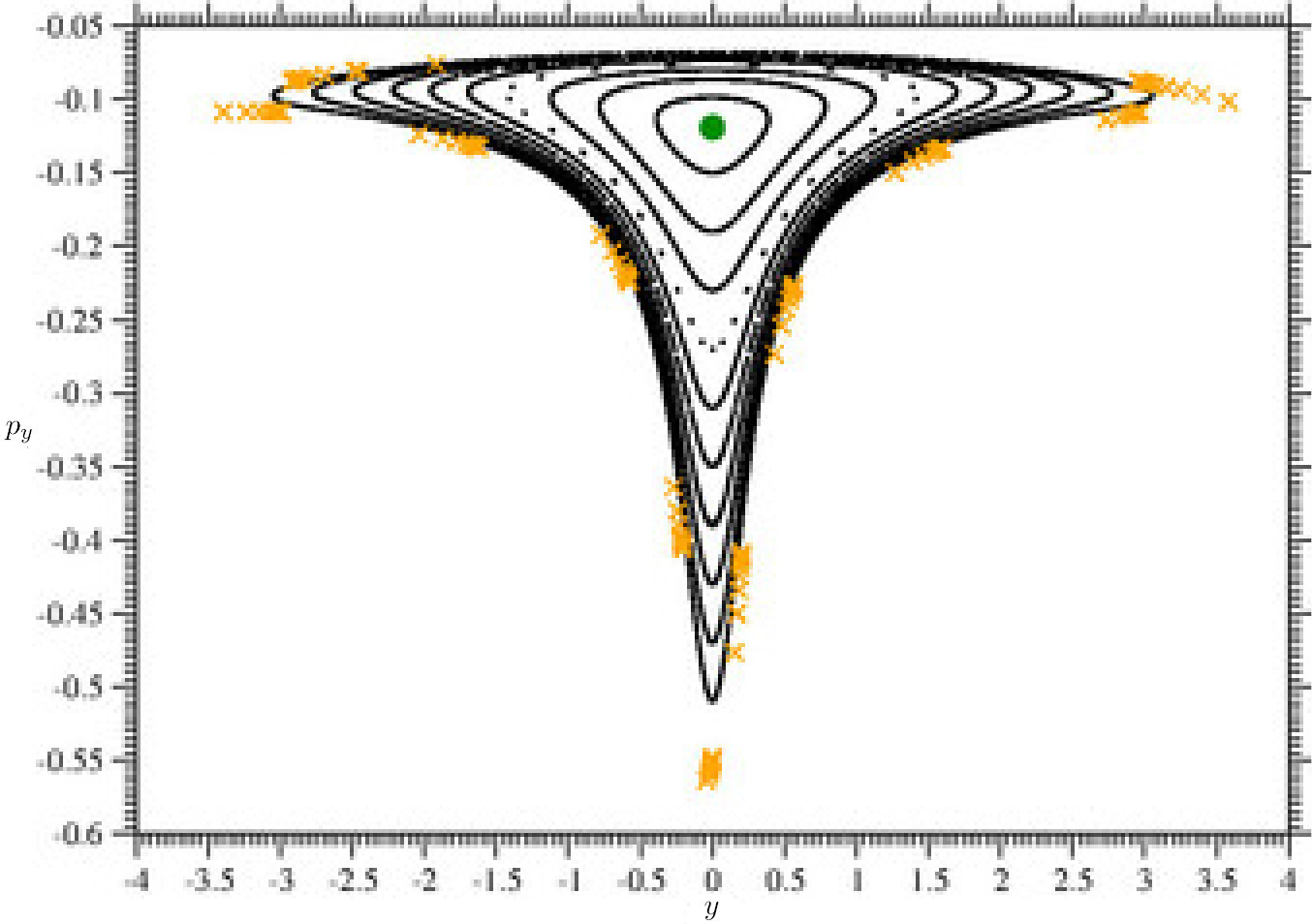}}
     {(d) $E_J = -0.041$}
     \\
     \subf{\includegraphics[scale=1.0 ]{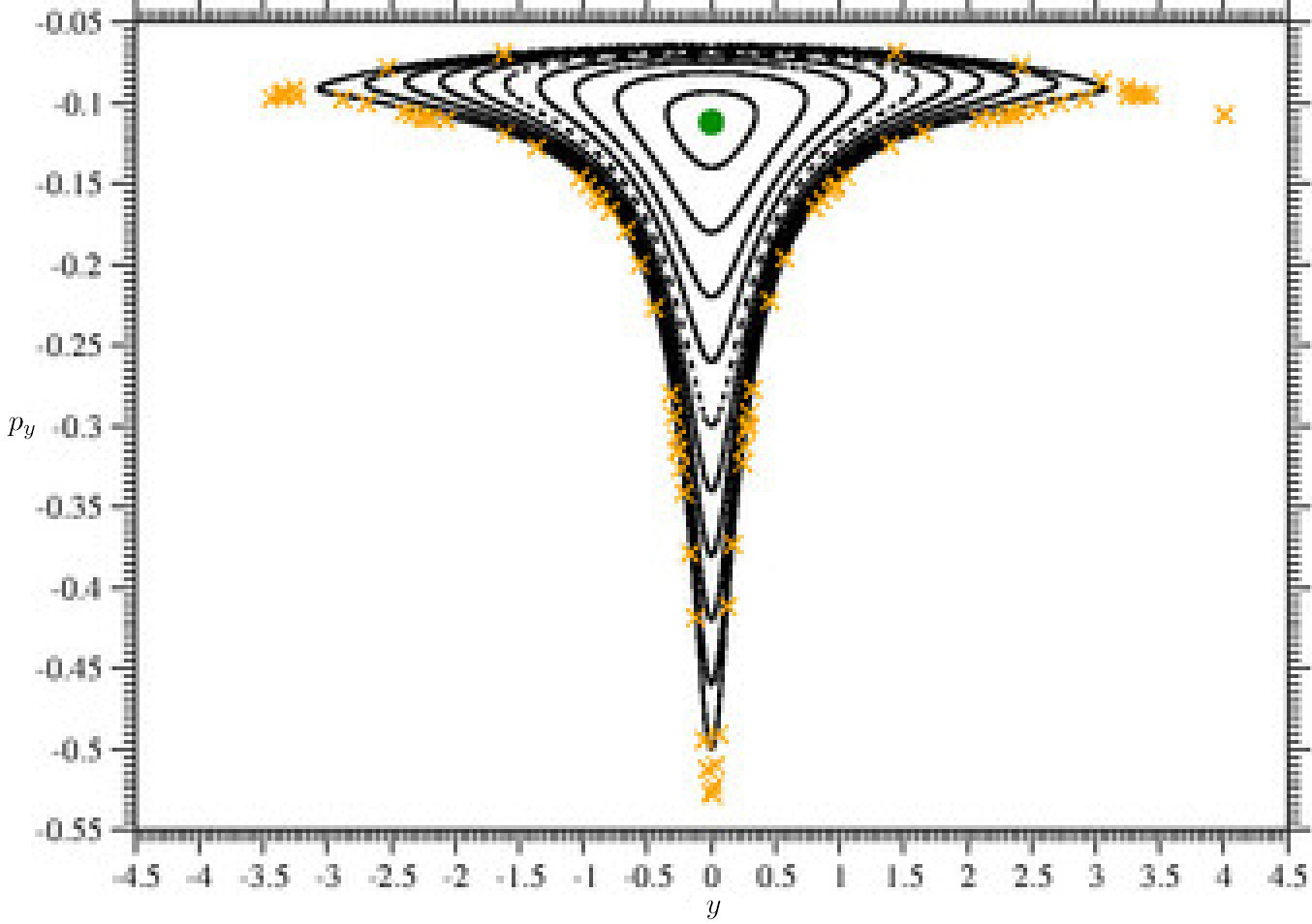}}
     {(e) $E_J = -0.04$}
    \end{tabular}
    \caption{ Projected Poincaré map of the NHIM $\mathcal{M}^3_{E_J}$  and its remnant with transient trajectories for different values of
        the Jacobi constant $E_J$ }
    \label{fig:4}
\end{figure}

At a Jacobi constant $E_J \approx -0.055$, $lh_2$ experiences a pitchfork bifurcation where it becomes tangentially hyperbolic and
splits off a pair of tangentially elliptic tilted loop orbits, which in the projected Poincaré map are the centres of KAM islands and
are again represented by blue dots in the figures. The Fig.\ref{fig:4}(a) shows the case of $E_J=-0.045$ where the islands around
the tilted loop orbits have already grown so large that they occupy approximately half of the area of the projected Poincaré map of NHIM $\mathcal{M}^3_{E_J}$.
Unlike $lh_1$ and as we know already from subsection 3.1 and Fig.\ref{fig:2}, $lh_2$ does
not run through a further pitchfork bifurcation and remains tangentially hyperbolic as long as it belongs to NHIM $\mathcal{M}^3_{E_J}$.
Note that for the present Jacobi constant, $lh_2$ has a higher normal instability than its tangential instability; therefore, it is still normally hyperbolic and forms part of NHIM $\mathcal{M}^3_{E_J}$.
In the projected Poincaré map, it is the outer part of a separatrix structure which it forms together with the separation line between the central island and the
tilted loop islands. Strictly speaking, this separatrix structure is again a fine chaos strip; however, so fine that it appears as lines under
the resolution of the graphics.

We are already in the interesting Jacobi constant region, therefore we increase $E_J$ in small steps and present in Fig.\ref{fig:4}(b)
the projected map of NHIM $\mathcal{M}^3_{E_J}$ for $E_J=-0.043$ even though the qualitative structure remains the same. We see that
in comparison with Fig.\ref{fig:4}(a) many invariant closed curves from
Fig.\ref{fig:4}(a) are broken into corresponding very fine secondary structures in Fig.\ref{fig:4}(b). This is a strong indication
that the system is already close to large-scale chaos and break of the NHIM $\mathcal{M}^3_{E_J}$.

In Fig.\ref{fig:4}(c), we proceed to $E_J=-0.042$ and see a very drastic change in comparison to Fig.\ref{fig:4}(b).
As we already know from Fig.\ref{fig:2}, for a Jacobi constant a little smaller than $E
_J = -0.042$, several important bifurcations of
the important periodic orbits in NHIM $\mathcal{M}^3_{E_J}$ occur. First, at $E_J \approx -0.0424$ $lh_2$ loses its
normal hyperbolicity and can no longer be a part of NHIM $\mathcal{M}^3_{E_J}$. Here, the tangential instability has
become larger than the normal instability. Second, at a Jacobi constant extremely close to this event, at $E_J \approx -0.04239$, $lh_2$
is destroyed in a saddle-centre bifurcation when it collides with another horizontal periodic orbit grown out of the
potential minimum $P_1$. Accordingly, the outer boundary and the separatrix structure of the NHIM $\mathcal{M}^3_{E_J}$
are lost. Third, at a Jacobi constant $E_J \approx -0.04232$, the tilted loop orbits from NHIM $\mathcal{M}^3_{E_J}$
collide with another pair of tilted loop orbits grown out of $P_1$ (see Fig.\ref{fig:2}). Thereby, the tilted loop
orbits together with their surrounding KAM tori are lost from NHIM $\mathcal{M}^3_{E_J}$. The only part of NHIM
$\mathcal{M}^3_{E_J}$ that persists this bifurcation process is the inner part of the central KAM tori around the
vertical Lyapunov orbit $lv_3$.

In the transient region where the tilted loop orbits have been destroyed, we find transient remnants of NHIM
$\mathcal{M}^3_{E_J}$ in analogy to what we saw before in Fig.\ref{fig:3}(f) for NHIM $\mathcal{M}^1_{E_J}$.
In Fig.\ref{fig:4}(c), the projection of the Poincaré map of one transient trajectory in the region is included by the orange crosses. We can imagine
that in the break of the 2-dimensional projected Poincaré map of the NHIM, lower-dimensional invariant subsets remain, and
this collection of lower-dimensional invariant sets can hold general trajectories in their neighbourhood at
least temporarily. Of course, these lower-dimensional unstable invariant sets also have stable and unstable
manifolds of corresponding lower dimension, which can hold very particular incoming trajectories forever.
But our numerical methods adapted to search for a 2-dimensional projected Poincaré map of the NHIM and their 3-dimensional stable manifolds do not
catch these lower-dimensional exceptional subsets. We only notice that general trajectories experience some time
delay in this region of the domain of the map.

In Fig.\ref{fig:4}(d), we display the projection of Poincaré map of NHIM $\mathcal{M}^3_{E_J}$ for $E_J= -0.041$. With increasing Jacobi constant,
the size of the central island shrinks slowly. Here in the outer transient region, we encountered something instructive.
We started a trajectory in the transient region close to the projected Poincaré map of NHIM $\mathcal{M}^3_{E_J}$  at the initial coordinates
$y=0$ and $p_y=-0.55$. This point happens to lie very close to the tangentially unstable central periodic point of a
secondary 5:11 resonance. First, the iterates lie close to this periodic orbit before they deviate strongly for higher
iterations and diffuse away. In total, we could follow this transient trajectory on the remnants of the NHIM
$\mathcal{M}^3_{E_J}$ for 125 iterations of the Poincaré map. The map of this transient trajectory is again plotted by orange crosses.
In Fig.\ref{fig:4}(e), we proceed to $E_J=-0.04$. Compared to the previous part, there is little difference.
Most interesting in this figure is the outer transient part. It is a single trajectory starting at $y=0$ and
$p_y = -0.538$. The algorithm could find 81 iterations of the Poincaré map, the last one is the point near $y=4$, $p_y = -0.107$.
We could not find a next iterate inside the region surrounding the NHIM
$\mathcal{M}^3_{E_J}$ to higher energies.

For still higher energies, the central island of the NHIM $\mathcal{M}^3_{E_J}$ shrinks slowly but remains.
For $E_J \approx -0.028$, $lv_3$ becomes tangentially unstable, but it remains normally hyperbolic even up
to positive energies. The continuation of NHIM $\mathcal{M}^3_{E_J}$ for such high energies is no longer relevant
for the theme of the present article; therefore, we do not follow it any longer.

\subsection{Bifurcation scenario of the NHIM $\mathcal{M}^2_{E_J}$}

\begin{figure} [h!]
\begin{center}
\includegraphics[scale=1.0 ]{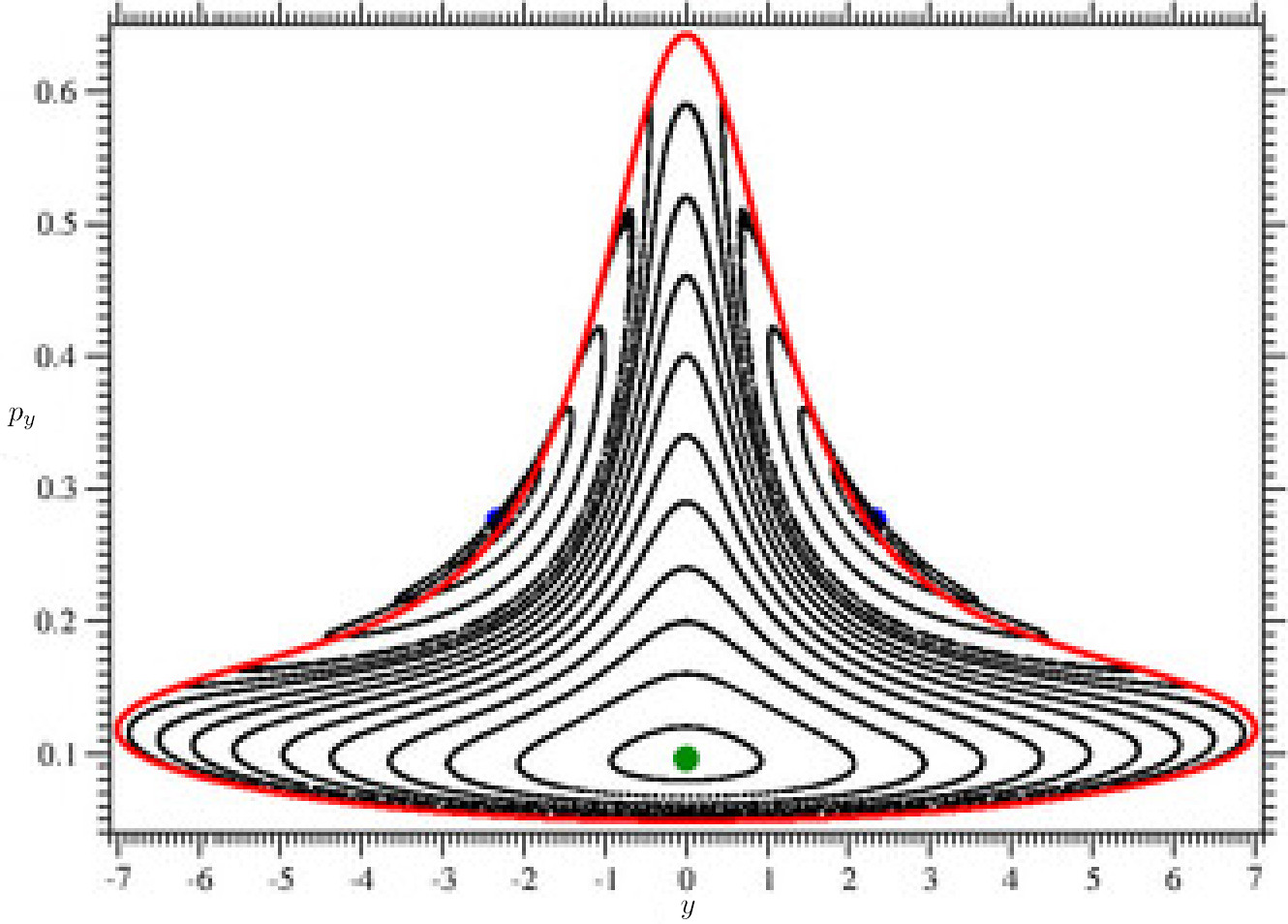}
\caption{Projected Poincaré map of the NHIM $\mathcal{M}^2_{E_J}$ for  Jacobi constant $E_J = -0.035$.}
\label{fig:5}   
\end{center}
\end{figure}

Finally, let us turn to NHIM $\mathcal{M}^2_{E_J}$. It is created at $E_{J2} \approx -0.05481$, and also this
NHIM starts with the internal dynamics of a 2-dof oscillator. Also, here the horizontal Lyapunov orbit $lh_2$
runs through a pitchfork where it splits off a pair of tilted loop orbits, it happens at $E_J \approx -0.046$,
see Fig.\ref{fig:2}. With increasing Jacobi constant, the KAM tori around the tilted loop orbits grow very
slowly, and we must reach rather high energies to obtain some interesting plot of the projected Poincaré map
of NHIM $\mathcal{M}^2_{E_J}$. In Fig.\ref{fig:5}, we show the
plot for $E_J = -0.035$. Of course, since its pitchfork bifurcation, $lh_2$ is unstable, and together with
the division between the central KAM island and the tilted loop islands in the Poincaré map, it forms a separatrix structure, which,
strictly speaking, is a very fine chaos strip. The apparent intersections between the various curves in the
tilted loop islands are a projection effect. Here, the projection of the curved NHIM $\mathcal{M}^2_{E_J}$
surface into the $ y$--$p_y $ plane is no longer 1:1; however, we always want to use the same projection for
all plots.

At $E_J \approx -0.032$ $lh_2$ first loses its normal hyperbolicity and immediately afterwards is destroyed
in a saddle-centre bifurcation when it collides with another horizontal periodic orbit growing out of the
potential minimum $P_2$. At $E_J \approx -0.02997$, the tilted loop orbits in NHIM $\mathcal{M}^2_{E_J}$ first
lose normal hyperbolicity and immediately afterwards are destroyed in saddle-centre bifurcations when they
collide with other tilted loop orbits grown out of $P_2$, see Fig.\ref{fig:2}. The vertical orbit $lv_2$
remains normally hyperbolic up to very high Jacobi constant, and accordingly, also a remnant of the NHIM
$\mathcal{M}^2_{E_J}$ around $lv_2$ remains. Qualitatively, the scenario of NHIM $\mathcal{M}^2_{E_J}$ is a
mirror image of the scenario of NHIM $\mathcal{M}^3_{E_J}$, even though the analogous individual events
happen at higher values of $E_J$. So we see in which sense a qualitative
mirror symmetry of the system is conserved, also when the symmetry between the masses is broken.

\subsection{Some comments on the NHIM bifurcation scenarios}

Important events happen in NHIM $\mathcal{M}^1_{E_J}$ and also in NHIM $\mathcal{M}^3_{E_J}$ in the same
relatively small Jacobi constant interval between $E =-0.043$ and $E_J=-0.04$. With
increasing Jacobi constant in NHIM $\mathcal{M}^3_{E_J}$, the loss of normal hyperbolicity happens before
the development of large-scale chaos in the internal dynamics. We see large-scale chaos in the surroundings
of the remnant parts of NHIM $\mathcal{M}^3_{E_J}$ only as a transient effect. In contrast, in the
internal dynamics of NHIM $\mathcal{M}^1_{E_J}$, we observe the development of large-scale chaos before
the beginning of any break of this NHIM. The difference is that NHIM $\mathcal{M}^3_{E_J}$ starts
its break because the normal instability is becoming small, whereas in NHIM $\mathcal{M}^1_{E_J}$,
the growth of the tangential instability is important for the beginning of the break.

When we compare the bifurcation scenario of NHIM $\mathcal{M}^1_{E_J}$ on one hand with the bifurcation
scenarios of NHIM $\mathcal{M}^2_{E_J}$ and NHIM $\mathcal{M}^3_{E_J}$ on the other hand, then we notice the
following qualitative difference. On NHIM $\mathcal{M}^1_{E_J}$, the horizontal Lyapunov orbit runs through
two consecutive pitchfork bifurcations, where in the second one, a pair of unstable tilted loop orbits is
created, while the horizontal Lyapunov orbit returns to tangential stability. This second pair of tilted
loop orbits turns vertical and is finally absorbed by the vertical Lyapunov orbit in an inverse pitchfork
bifurcation. Thereby, the vertical Lyapunov orbit becomes tangentially unstable, and this event triggers
the creation of large-scale chaos inside of NHIM $\mathcal{M}^1_{E_J}$.

In contrast, in NHIM $\mathcal{M}^2_{E_J}$ and in NHIM $\mathcal{M}^3_{E_J}$, tangentially unstable tilted
loop orbits are never created, which has two consequences. First, the horizontal Lyapunov orbit remains
tangentially unstable, which favours the beginning of the break of these NHIMs from the
separatrix structure, which contains $lh_2$ or $lh_3$, respectively. Second, the vertical Lyapunov orbit
remains tangentially stable, which favours the survival of the central KAM island of the Poincaré map of these NHIMs up to very high Jacobi constant.

\begin{figure}[h!]
\begin{center}
\includegraphics[scale=1.0 ]{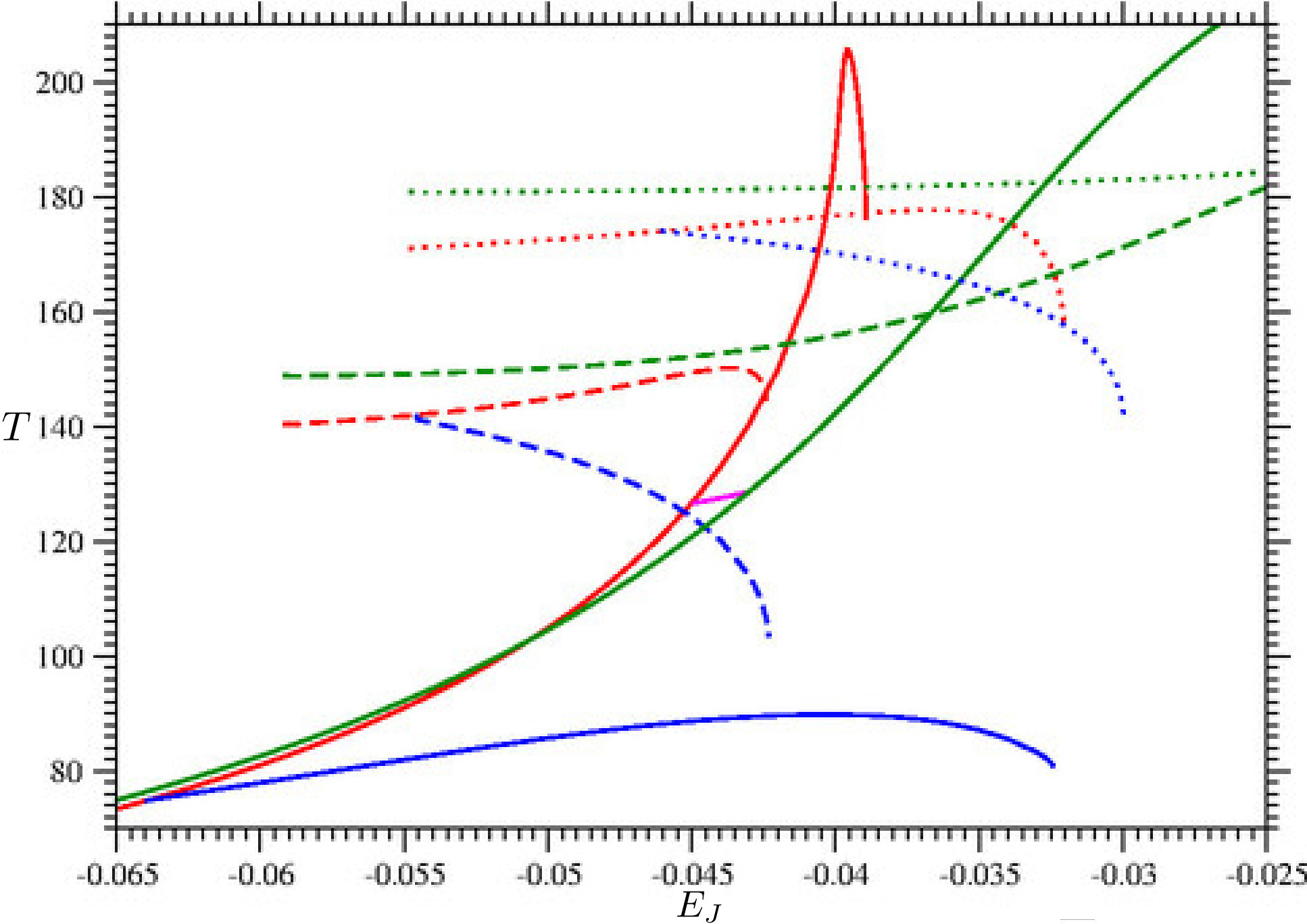}
\caption{Periods of Lyapunov orbits and tilted loop orbits as a function of the Jacobi constant $E_J$.}
\label{fig:6}
\end{center}
\end{figure}

We made a surprising observation of the connection between these properties and the behaviour of the
time periods of the Lyapunov orbits as a function of the Jacobi constant. In Fig.\ref{fig:6}, we plot
these time periods as a function of the Jacobi constant. As before, the red curves belong  to the
horizontal Lyapunov orbits, and the green curves belong to the vertical Lyapunov orbits. The blue curves belong to the tangentially
stable tilted loop orbits. And the magenta curve belongs to the tangentially unstable tilted loop orbits. The solid curves belong to
NHIM $\mathcal{M}^1_{E_J}$, the dotted curves belong to NHIM $\mathcal{M}^2_{E_J}$ and the broken
curves belong to NHIM $\mathcal{M}^3_{E_J}$. In NHIM $\mathcal{M}^1_{E_J}$, the periods of the horizontal and 
the vertical Lyapunov orbit cross each other; one pitchfork bifurcation happens energetically before the
crossing, and the other one happens after the crossing. In NHIM $\mathcal{M}^2_{E_J}$ and NHIM $\mathcal{M}^3_{E_J}$,
we do not find any analogous crossing of the horizontal and vertical periods.

Now the reader might ask whether this observation is a strange feature of the particular example or whether it is common.
Interestingly, the experience with other systems suggests that this behaviour is common. In the system of a test particle in a
barred galaxy \cite{zj4} and for the dynamics of a test particle in a double-barred galaxy \cite{zj2}, there are two consecutive
pitchfork bifurcations creating first a pair of tangentially stable tilted loop orbits and later a pair of tangentially unstable
tilted loop orbits. Also, a crossing of the periods of the horizontal and the vertical Lyapunov orbits occurs between
these two pitchfork bifurcations. However, in these two examples, the tilted loop orbits come out of the vertical Lyapunov orbit.
On the other hand, in the system of a test particle in the gravitational field of an open star cluster \cite{zj5}, there is
no crossing of periods, and there is only a single pitchfork bifurcation creating a pair of tangentially stable tilted loop orbits.

In a classical model for a hydrogen atom in a rotating external field \cite{jw},  the scenario is as follows. First the horizontal
Lyapunov orbit splits off a pair of tangentially stable tilted loop orbits. Later, there is a crossing of the periods of the
horizontal and vertical Lyapunov orbits. However, at a Jacobi constant a little higher, the horizontal Lyapunov orbit collides with the
singularity of the Coulomb system at the origin, and this event throws the scenario out of its usual path. At a Jacobi constant still
a little higher, the horizontal Lyapunov orbit is absorbed by an orbit coming out of the central potential hole in an inverse
period doubling bifurcation. So the horizontal Lyapunov orbit ends its existence before it has the opportunity for a further
pitchfork bifurcation.

The conclusion of this subsection is the following. We have seen the creation of tangentially unstable tilted loop orbits only
in cases where the periods of the horizontal and the vertical Lyapunov orbits cross. The system of the two dwarf galaxies is
a nice example where we see in the same system the case with crossing and the case without crossing. Certainly, the crossing of
the periods indicates some kind of 1:1 resonance between the horizontal and the vertical motion. And such a resonance is needed
for the creation of tilted loop orbits out of the Lyapunov orbits. Unfortunately, so far, we are not yet able to give a detailed
explanation of all these connections.

\newpage

\section{Heteroclinic connections and the appearance of large-scale chaos}

\begin{figure}[h!]
\begin{center}
   \includegraphics[scale=1.0 ]{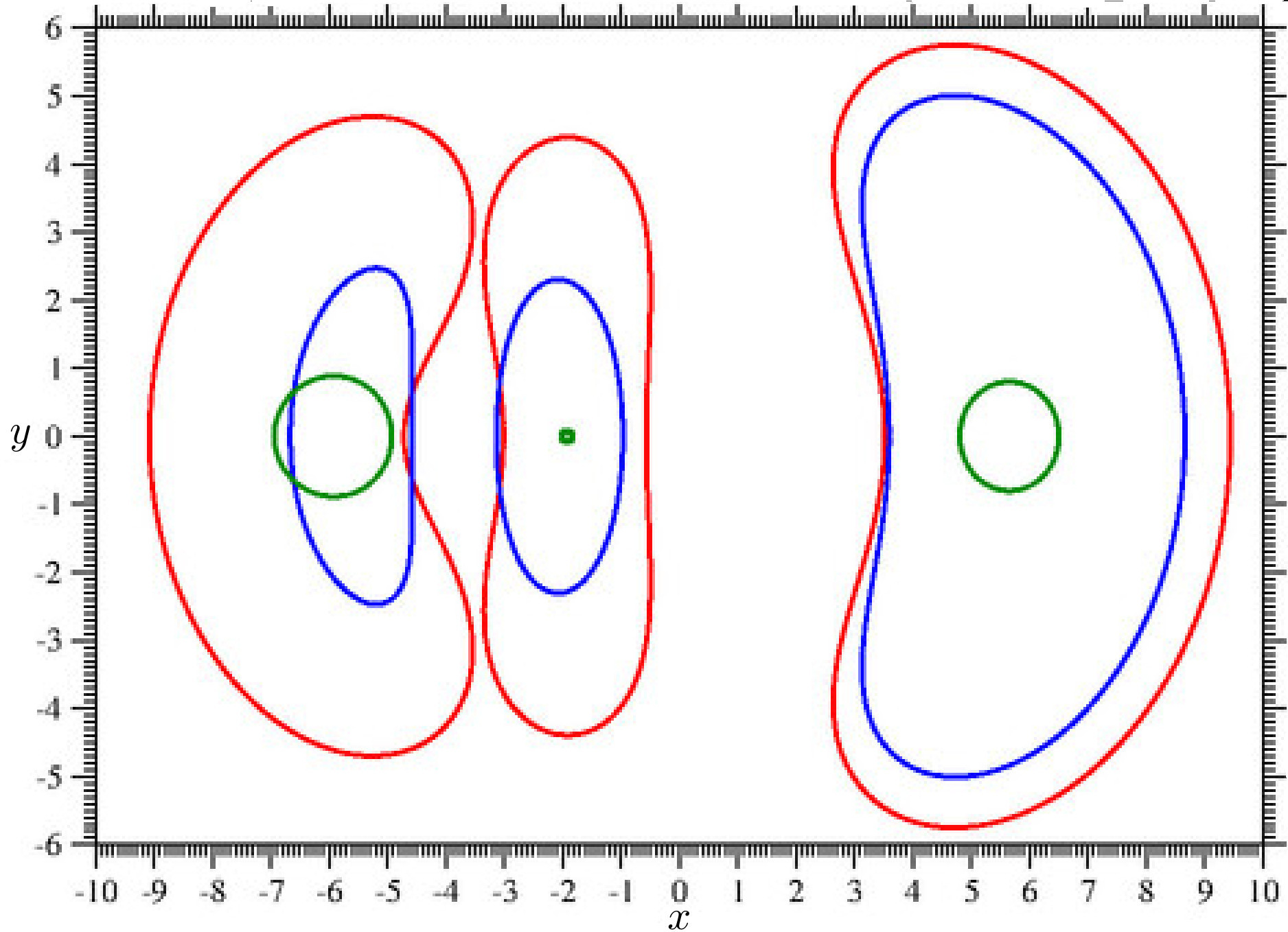}
\caption{Projection of the simplest and most important periodic
orbits at $E_J = -0.043$ into the horizontal plane of the position space.
The red curves are horizontal Lyapunov orbits, the green curves are vertical Lyapunov orbits, 
and the blue curves are tangentially stable tilted loop orbits.}
\label{fig:7}
\end{center}
\end{figure}

Now we relate the NHIM bifurcation scenarios seen in section 3 to 
some particularly interesting classes of trajectories and their properties, namely, to heteroclinic trajectories between
the 3 NHIMs.
We do it for the Jacobi constant  values  $E_J =-0.043$, $-0.042$ and $-0.04$, i.e. exactly for the Jacobi
constant range where the most interesting events of the whole bifurcation scenario happen. Because of the easier
pictorial presentation, we begin with horizontal trajectories. The only horizontal trajectories within the NHIMs are
the respective horizontal Lyapunov orbits, which have already served as boundaries of the NHIM representations in the
plots of the projected Poincaré map. Fig.\ref{fig:7} shows the three horizontal Lyapunov orbits $lh_1$, $lh_2$ and $lh_3$
(plotted as red lines) in the horizontal plane of the position space for $E=-0.043$. These three orbits are identified in
the plot by their order along the $x$ axis. We know already from Fig.\ref{fig:2} that this Jacobi constant value is very
close to the highest Jacobi constant where all the important fundamental periodic orbits still exist.  The figure contains,
in addition, the projection into the horizontal plane of the vertical Lyapunov orbits (dark green curves) and of the tilted
loop orbits (blue curves). The orbit $lu_1$ disappears already exactly at this Jacobi constant when it collides with $lv_1$.
Therefore, it looks very similar to $lv_1$ at this Jacobi constant, and it is not included in the figure because it
would be extremely difficult to distinguish $lu_1$ and $lv_1$.
Note that $lh_3$ and $lh_1$ come very close to each other, while $lh_2$ keeps a greater distance from $lh_1$.
The two tilted loop orbits of each NHIM have identical projections into the horizontal plane.

The trace of any one of these periodic orbits is symmetric under $y$ reflection; the $y$ reflection only reverses the time 
orientation of the orbits. This is a consequence of the symmetry from Eq.6 under which the whole dynamics is invariant.

\subsection{Heteroclinic trajectories and transient dynamics associated with the NHIMs break}

\begin{figure}[h!]
    \centering
    \begin{tabular}{c c}   
     \subf{\includegraphics[scale=1.0]{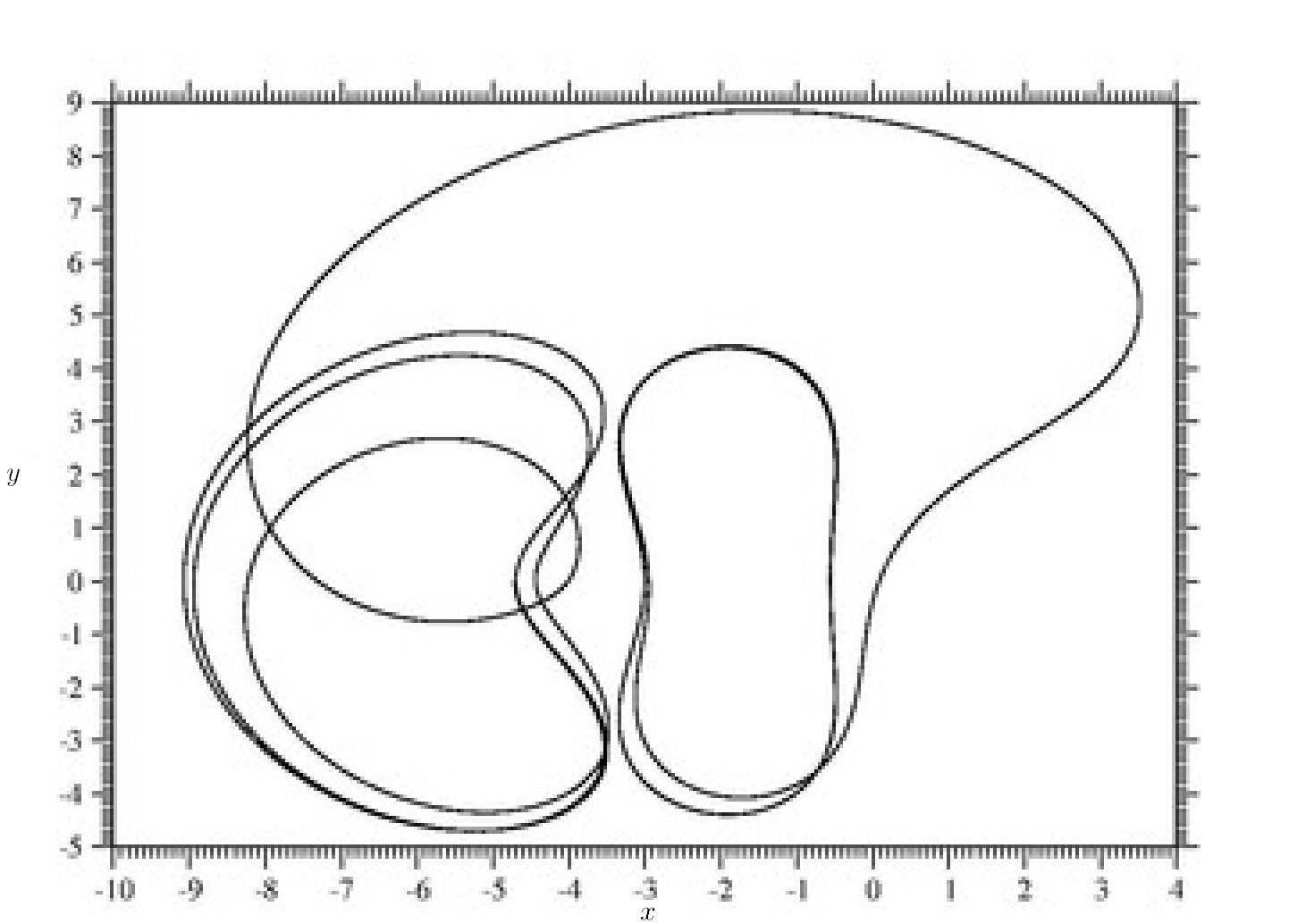}}
     {(a)}
      &
     \subf{\includegraphics[scale=1.0 ]{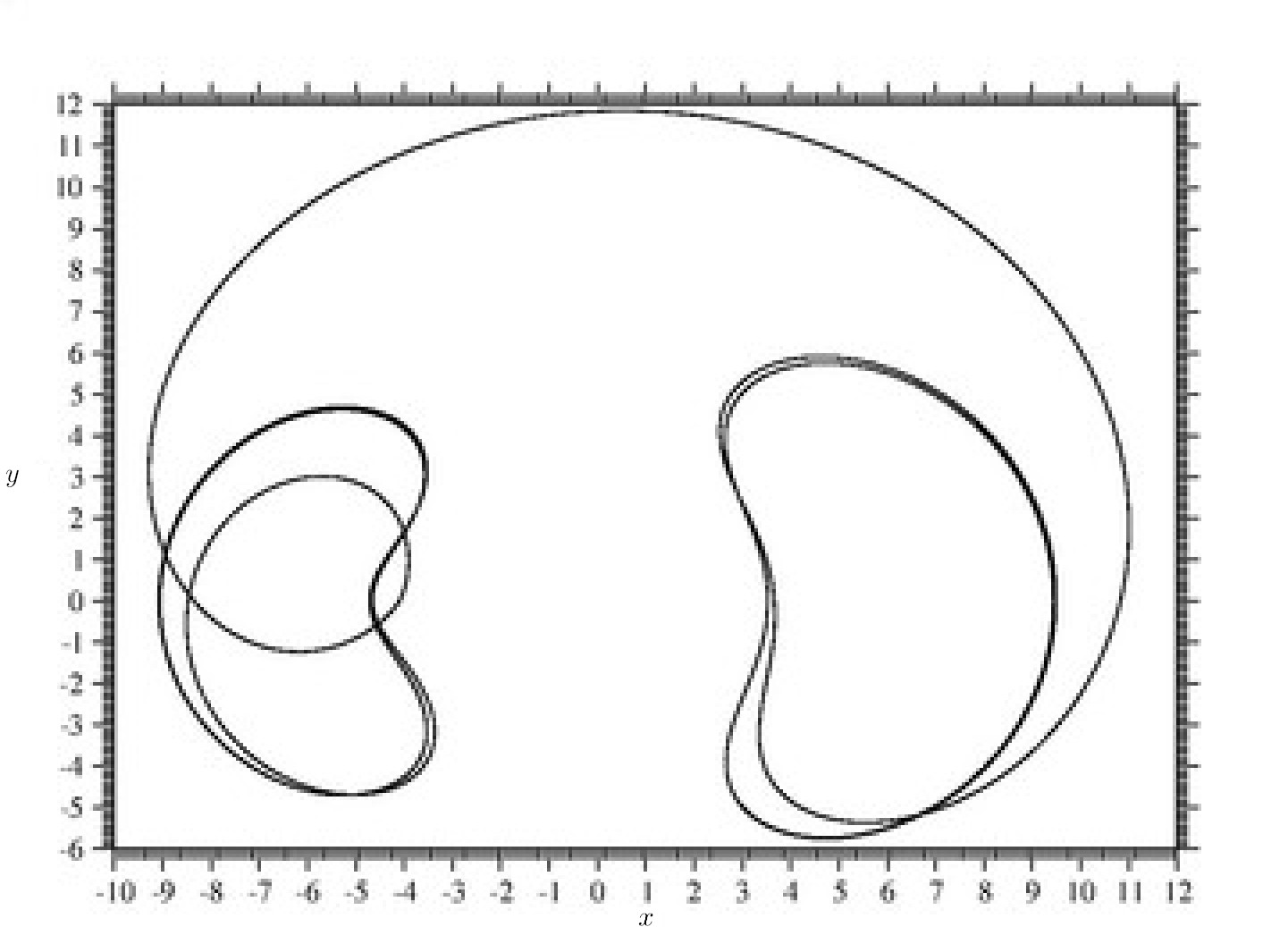}}
     {(b) }
     \\
     \subf{\includegraphics[scale=1.0 ]{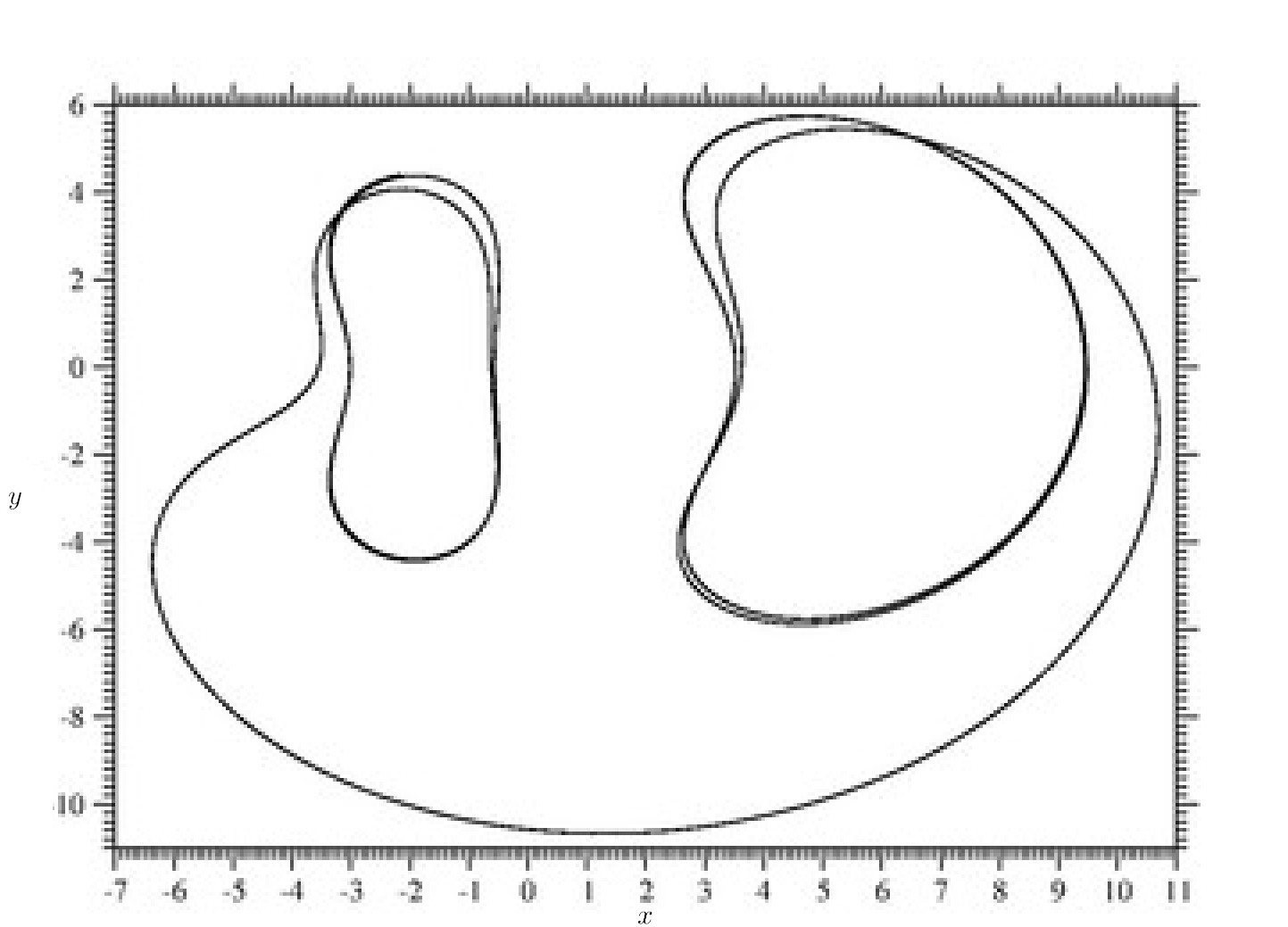}}
     {(c) }
     &
     \subf{\includegraphics[scale=1.0 ]{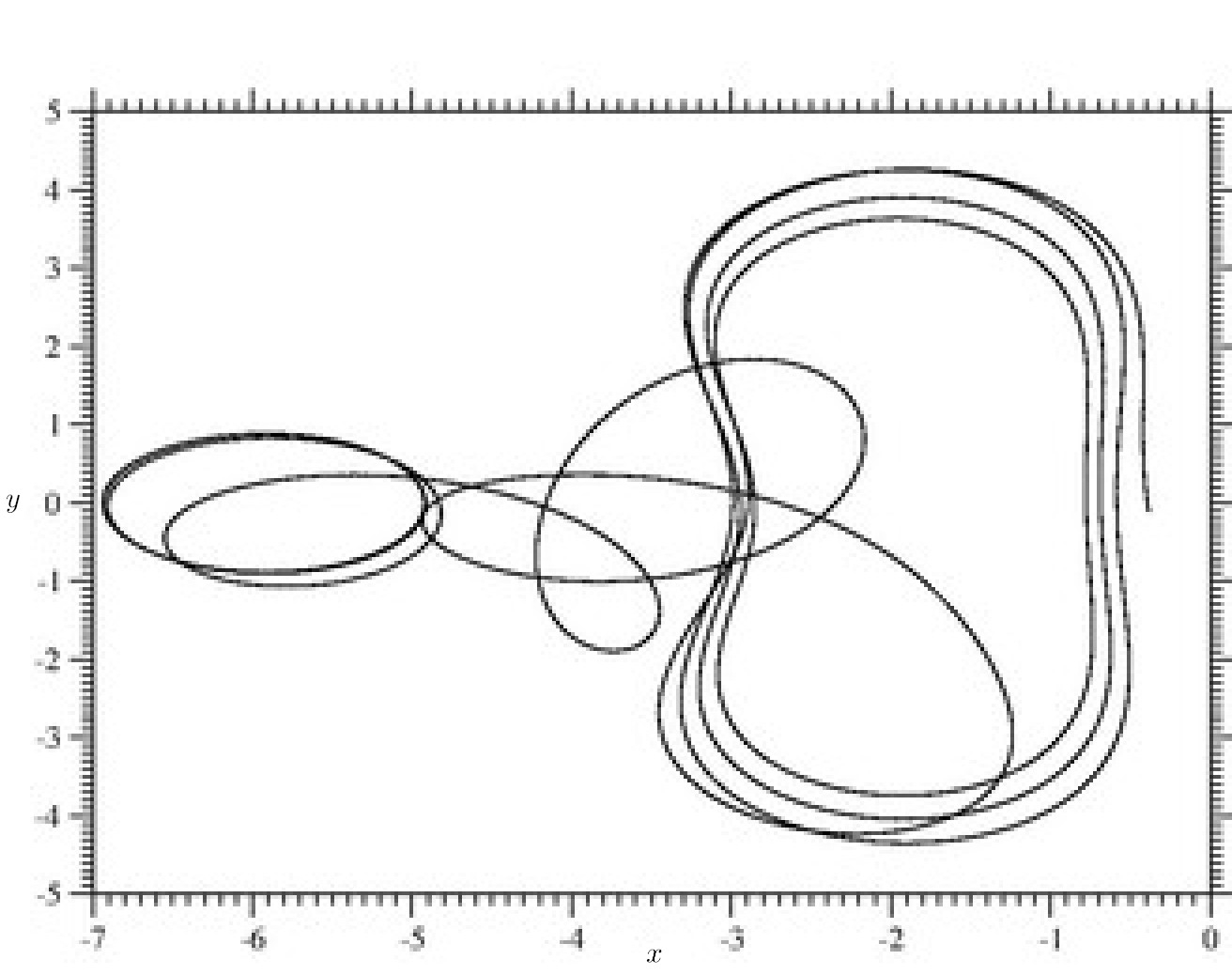}}
     {(d)}
     \\
     \subf{\includegraphics[scale=1.0 ]{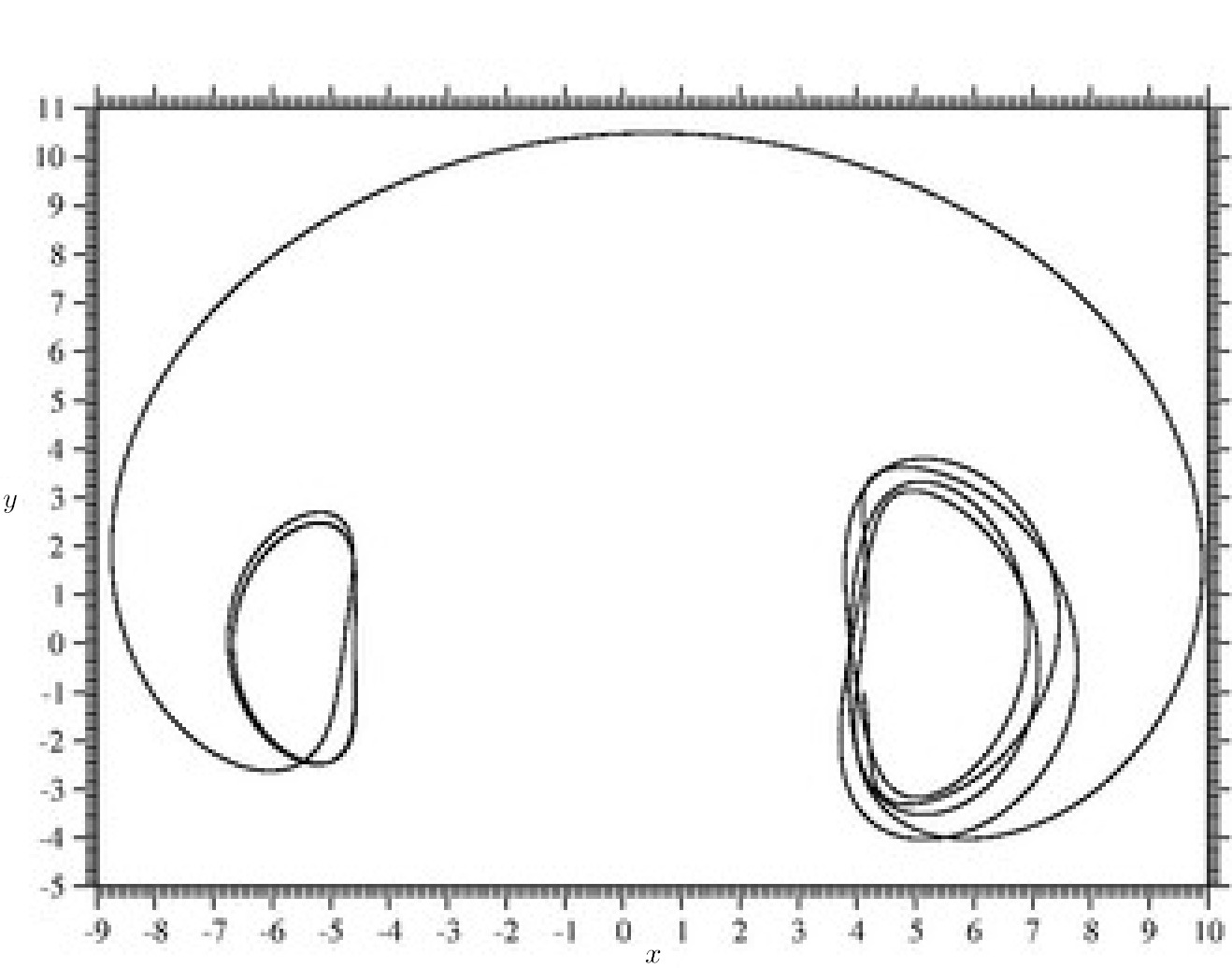}}
     {(e) }
    \end{tabular}
    \caption{ Heteroclinic trajectories between various substructures of the NHIMs at $E_J = -0.043$.
(a) : A horizontal heteroclinic trajectory going from $lh_3$ to $lh_1$.
(b) : A horizontal heteroclinic trajectory going from $lh_3$ to $lh_2$.
(c) : A horizontal heteroclinic trajectory going from $lh_2$ to $lh_1$.
(d) : Projection into the horizontal plane of a heteroclinic trajectory
      going from $lv_3$ to the outer part of the tilted loop island of NHIM $\mathcal{M}^1_{E_J}$.
(e) : Projection into the horizontal plane of a heteroclinic trajectory
      going from $lt_3$ to the outer part of the central island of NHIM $\mathcal{M}^2_{E_J}$.
}
    \label{fig:8}
\end{figure}

First, we have a look at heteroclinic orbits between the horizontal Lyapunov orbits. They are the only horizontal
heteroclinic orbits between the various NHIMs. In Fig.\ref{fig:8}(a), we show a typical example of a heteroclinic
trajectory which starts on $lh_3$ and ends on $lh_1$. In Fig.\ref{fig:8}(b), we show a heteroclinic trajectory
going from $lh_3$ to $lh_2$ and Fig.\ref{fig:8}(c) presents a heteroclinic trajectory going from $lh_2$ to $lh_1$.
Of course, there is an infinity of further horizontal heteroclinic trajectories between the various horizontal
Lyapunov orbits perform all kinds of loops in the regions of the NHIMs before settling down to one particular
Lyapunov orbit in the past and  in the future.

We also show two examples of heteroclinic trajectories going out of the horizontal plane. However, in the figures,
we plot their projections into the horizontal plane. Fig.\ref{fig:8}(d) presents a trajectory starting on $lv_3$
and ending on the outer part of the tilted loop island of NHIM $\mathcal{M}^1_{E_J}$ close to the separatrix.
Fig. \ref{fig:8}(e) presents a trajectory starting on $lt_3$ and ending on the outer part of the
central island of NHIM $\mathcal{M}^2_{E_J}$.

\begin{figure}[h!]
    \centering
    \begin{tabular}{c c}   
     \subf{\includegraphics[scale=1.0 ]{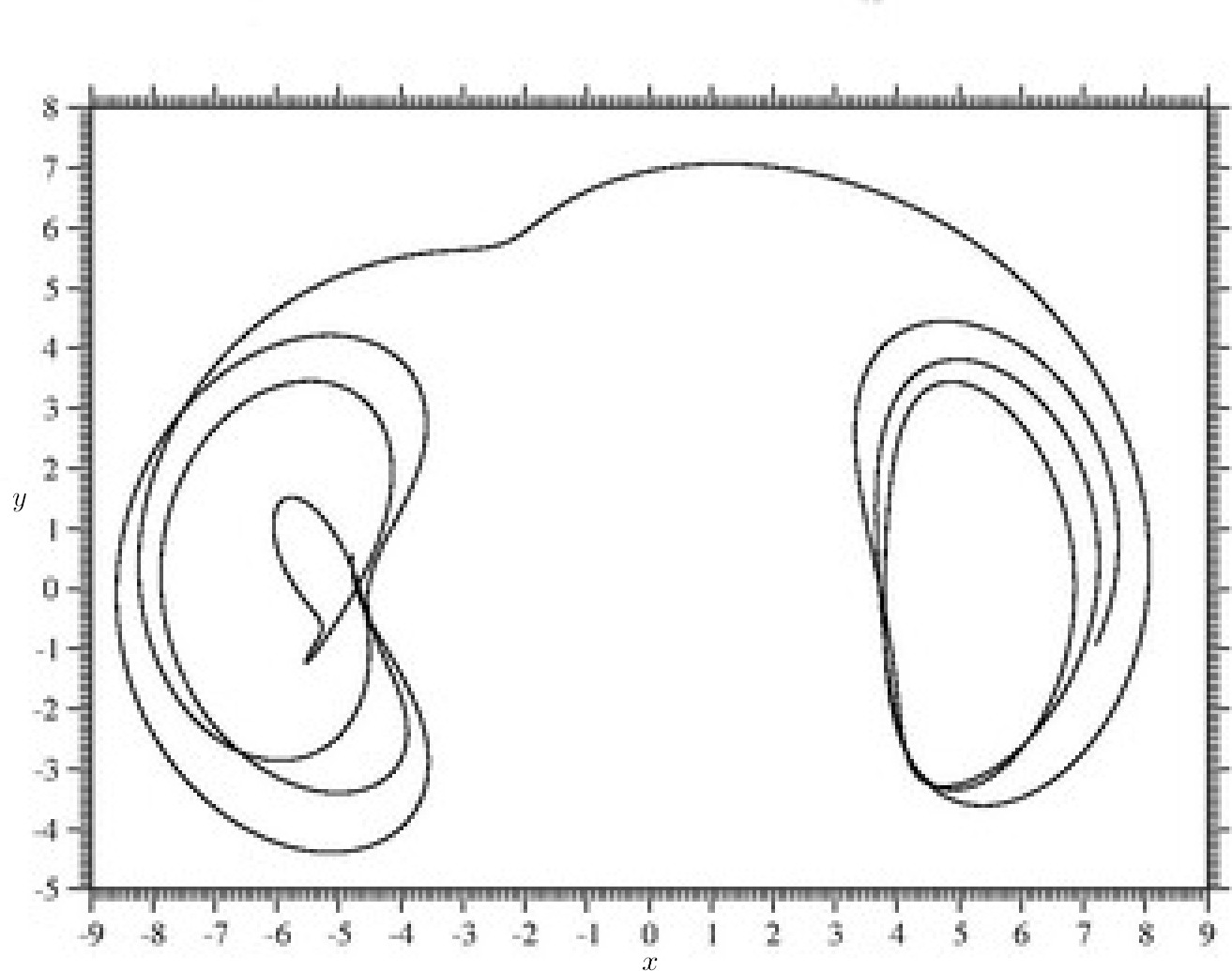}}
     {(a)}
      &
     \subf{\includegraphics[scale=1.0 ]{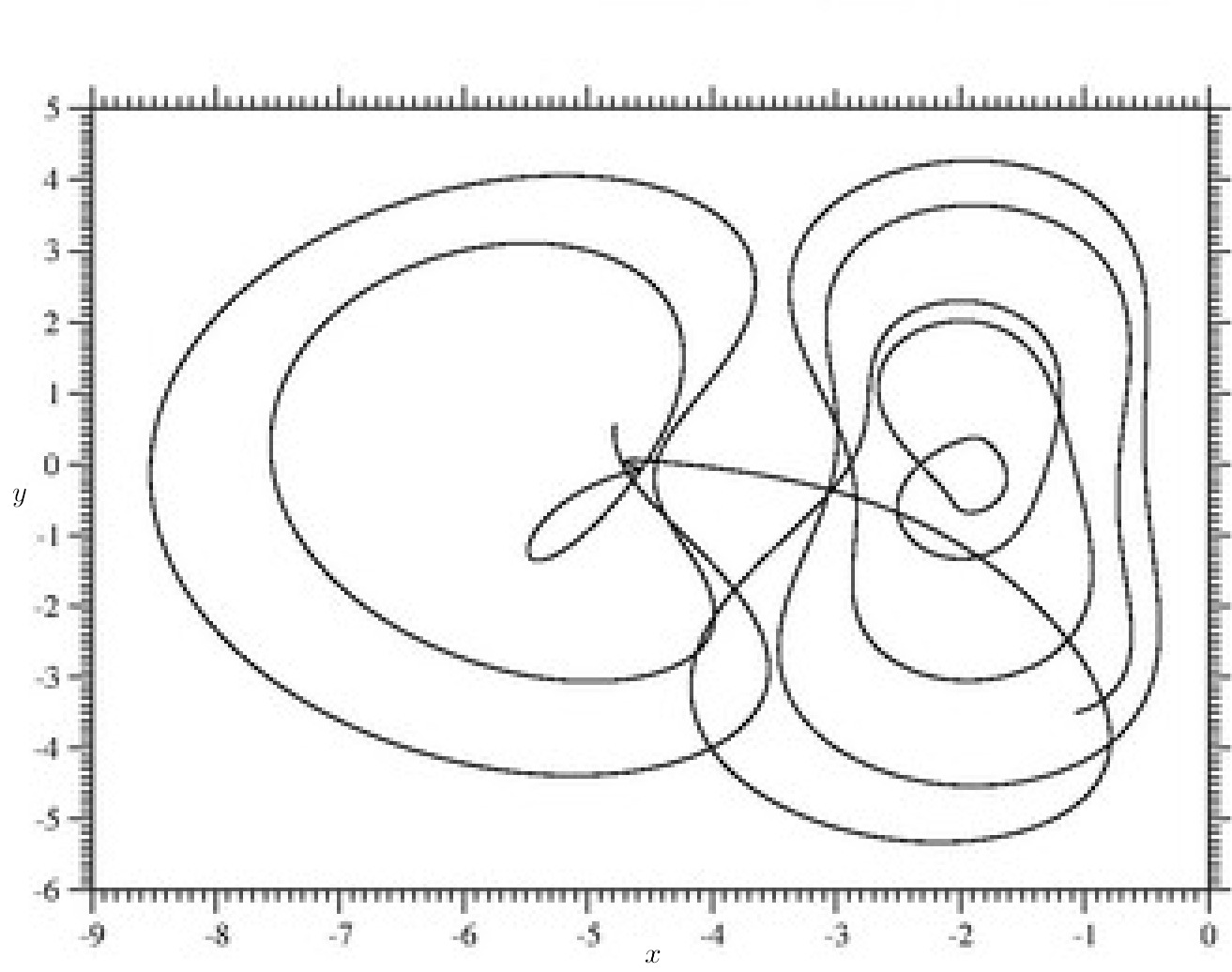}}
     {(b)}
     \\
     \subf{\includegraphics[scale=1.0 ]{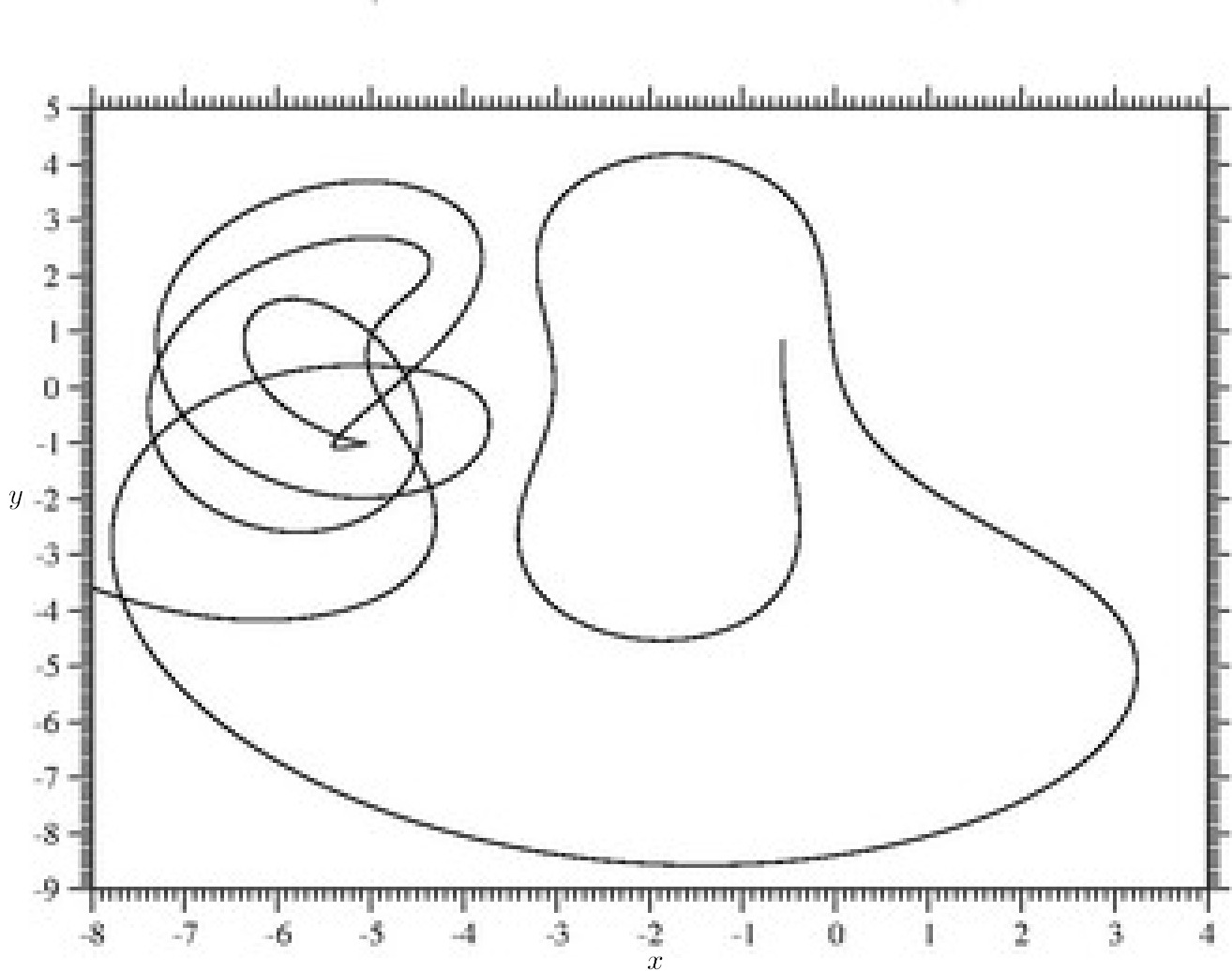}}
     {(c)}
     &
     \subf{\includegraphics[scale=1.0 ]{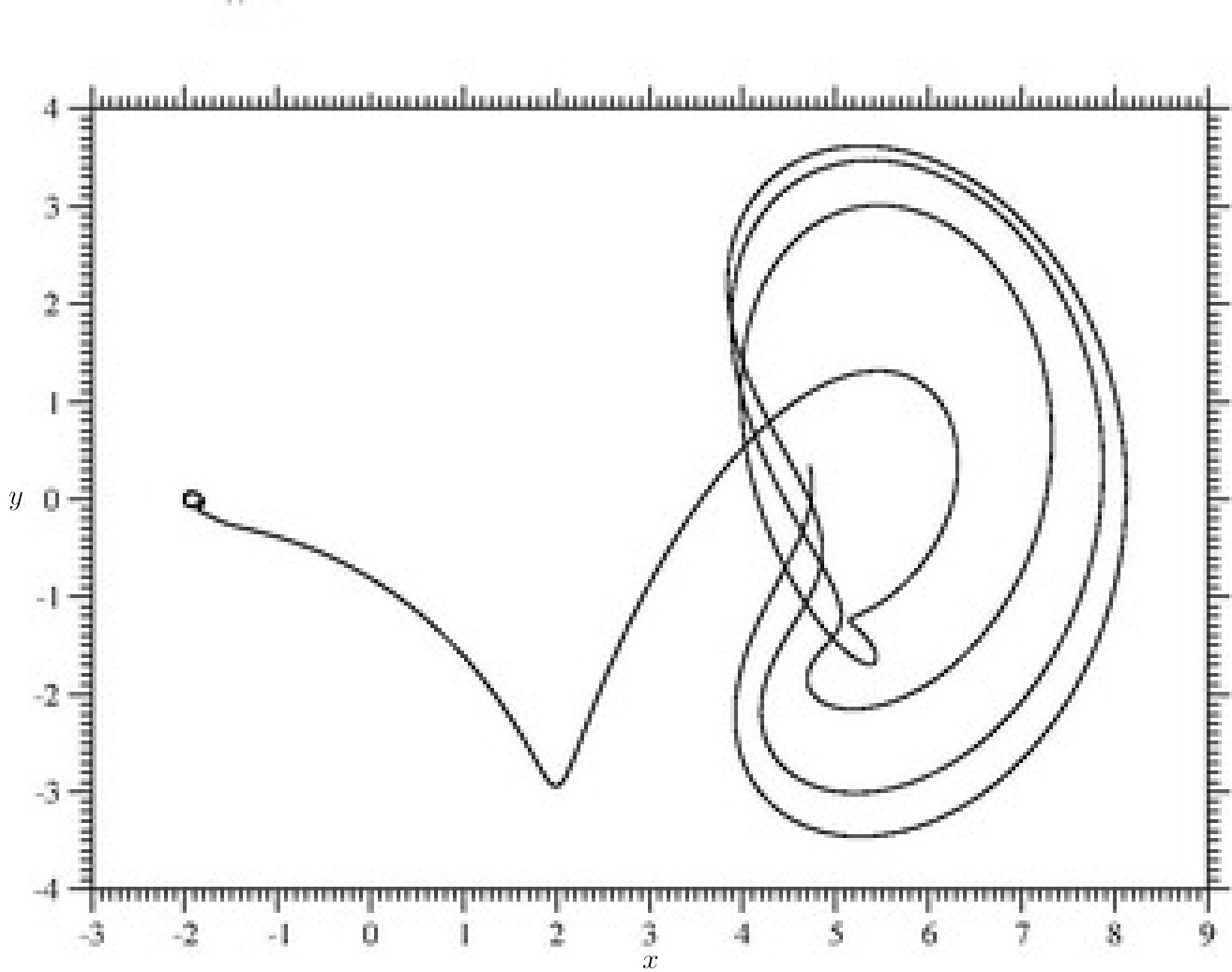}}
     {(d) }

    \end{tabular}
    \caption{ Projection into the horizontal plane of four trajectories between the neighbourhoods of NHIMs at $E_J = -0.042$.
Part (a): A trajectory from the transient part close to NHIM $\mathcal{M}^3_{E_J}$ to NHIM $\mathcal{M}^2_{E_J}$.
part (b): A trajectory from the transient part close to NHIM $\mathcal{M}^3_{E_J}$ to NHIM $\mathcal{M}^1_{E_J}$.
part (c): A trajectory from the large-scale chaotic region of NHIM $\mathcal{M}^1_{E_J}$ to NHIM $\mathcal{M}^3_{E_J}$.
part (d): A heteroclinic from the central region of NHIM $\mathcal{M}^1_{E_J}$, close to the trajectory from $lv_1$, 
to NHIM $\mathcal{M}^2_{E_J}$. }
    \label{fig:9}
\end{figure}

Now we increase the Jacobi constant a little to $E_J=-0.042$. We know already that for this Jacobi constant, the periodic
orbit $lh_3$ has already disappeared in a saddle-centre bifurcation included in Fig.\ref{fig:2}. Accordingly,
there is no longer any horizontal trajectory inside the  NHIM $\mathcal{M}^3_{E_J}$ and therefore
also no horizontal heteroclinic trajectory starting or ending on NHIM $\mathcal{M}^3_{E_J}$. However, we have the
transient chaotic region close to the NHIM $\mathcal{M}^3_{E_J}$ on the plot of Fig.\ref{fig:4}(c). And some of these
transient trajectories close to NHIM $\mathcal{M}^3_{E_J}$ have a projection into the horizontal plane
which qualitatively looks similar to $lh_3$ at smaller energies, i.e. it looks as if there would be a logical
continuation of $lh_3$ to higher energies. From the neighbourhood of such transient trajectories, other
trajectories can go to other NHIMs. Fig.\ref{fig:9}(a) shows the projection into the horizontal plane of the
position space of a typical trajectory going from the transient region close to NHIM $\mathcal{M}^3_{E_J}$ to NHIM
$\mathcal{M}^2_{E_J}$. Its initial point lies very close to the 26th Poincaré map iteration of the transient trajectory shown
in Fig.\ref{fig:4}(c), its coordinates in this figure are $y \approx 0.57, p_y \approx -0.39$. On the NHIM
$\mathcal{M}^2_{E_J}$, this transient chaotic trajectory approaches the outer part
of the central island in NHIM $\mathcal{M}^2_{E_J}$.

Fig.\ref{fig:9}(b) shows the projection into the horizontal plane of the position space of a trajectory going from
the transient region close to NHIM $\mathcal{M}^3_{E_J}$ to NHIM $\mathcal{M}^1_{E_J}$. The initial point of this trajectory
lies extremely close to the initial point of the trajectory shown in Fig.\ref{fig:9}(a). Compare in the figures the
initial point and the first loop of these two trajectories in the region of NHIM $\mathcal{M}^3_{E_J}$. However, in the long
run, these two trajectories are completely different. This demonstrates the extreme dependence on the initial conditions of
trajectories starting in the chaotic region, a typical indication of chaos. This trajectory approaches the
NHIM $\mathcal{M}^1_{E_J}$ in the region between the moderate-sized chaos strip created around the stable and unstable
manifolds of the central hyperbolic fixed point corresponding to the vertical Lyapunov orbit $lv_1$ and the outermost
layer of the tilted loop islands, which is already broken into secondary structures and chaos strips. It establishes a
connection between the transient chaotic region around NHIM $\mathcal{M}^3_{E_J}$ and the part of NHIM
$\mathcal{M}^1_{E_J}$, which is just in the process of turning chaotic at this stage of the bifurcation scenario.

Fig.\ref{fig:9}(c) shows a trajectory starting in the chaotic region of NHIM $\mathcal{M}^1_{E_J}$ and coming close
to the transient chaotic region close to NHIM $\mathcal{M}^3_{E_J}$.
Fig.\ref{fig:9}(d) shows a trajectory starting close to $lv_1$, i.e. in the centre of the chaotic region of NHIM
$\mathcal{M}^1_{E_J}$ and approaching NHIM $\mathcal{M}^2_{E_J}$.

\begin{figure}[h!]
    \centering
    \begin{tabular}{c c}   
     \subf{\includegraphics[scale=1.0 ]{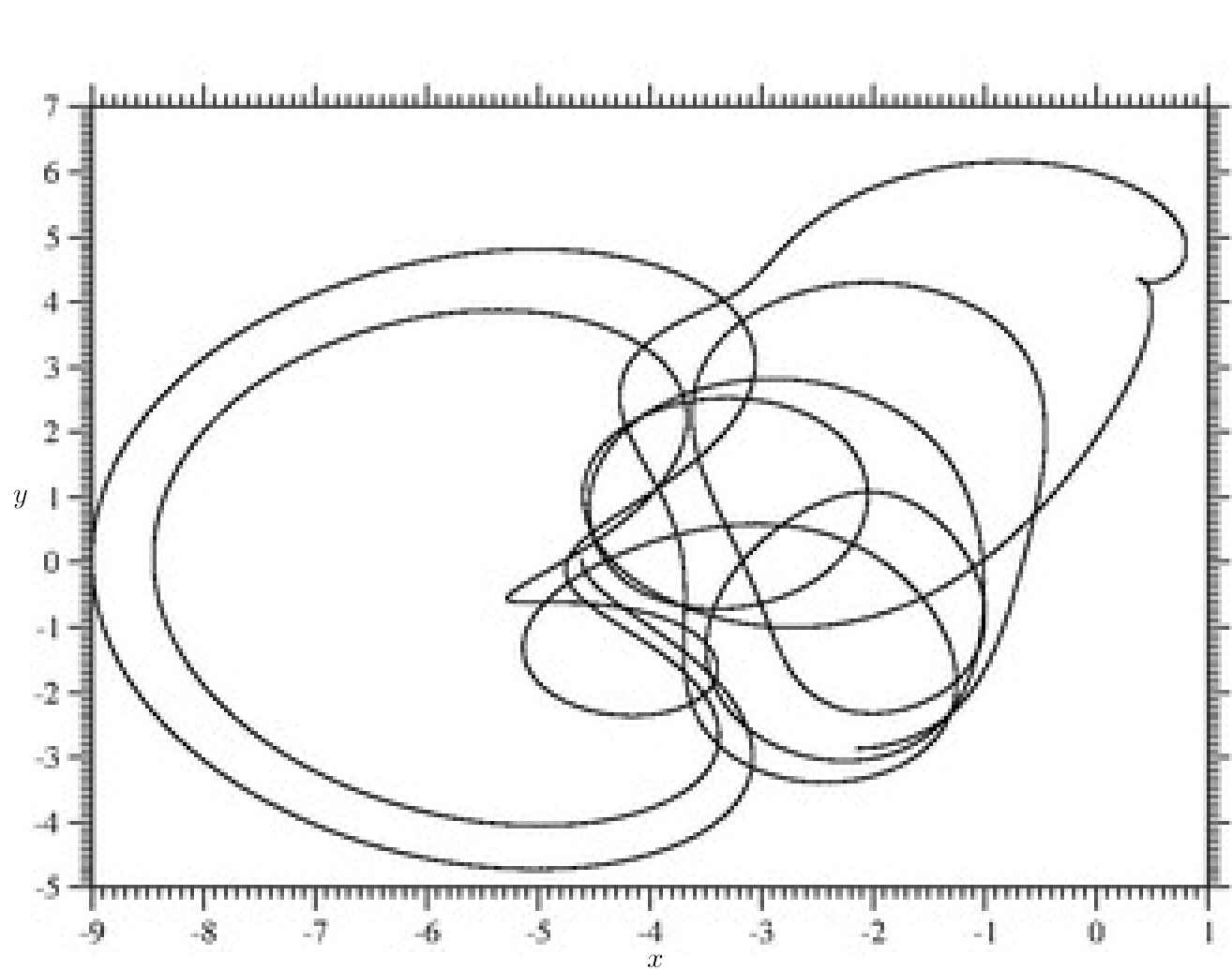}}
     {(a)  }
      &
     \subf{\includegraphics[scale=1.0 ]{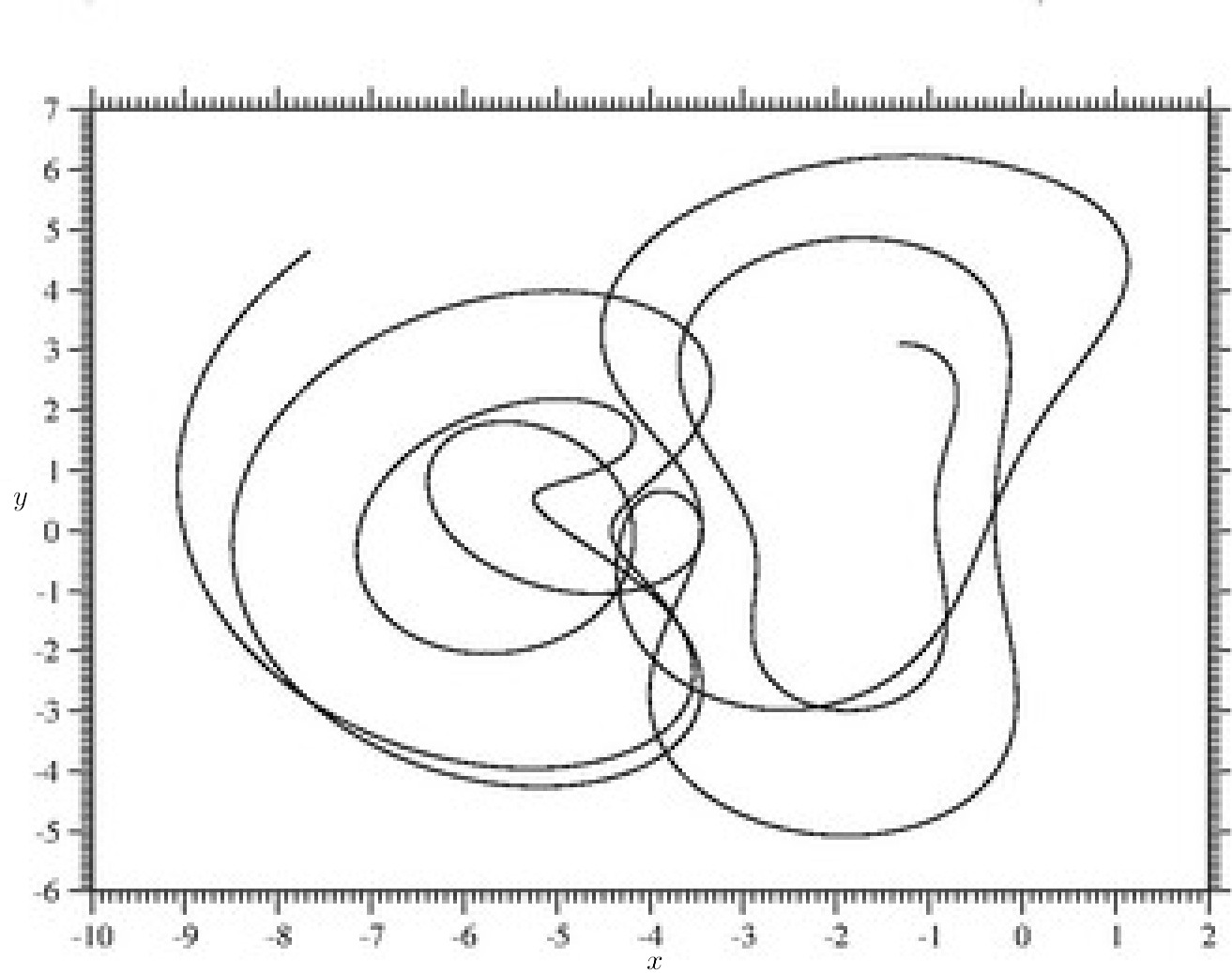}}
     {(b)  }
     \\

    \end{tabular}
    \caption{ Projection into the horizontal plane of two trajectories at $E_J = -0.04$.
Part (a): A trajectory going from the transient region close to NHIM $\mathcal{M}^3_{E_J}$ to NHIM $\mathcal{M}^1_{E_J}$.
Part (b): A trajectory going from the transient region close to NHIM $\mathcal{M}^1_{E_J}$ to NHIM $\mathcal{M}^3_{E_J}$.}
    \label{fig:10}
\end{figure}

Next, let us consider $E_J= -0.04$ where NHIM $\mathcal{M}^3_{E_J}$ and NHIM $\mathcal{M}^1_{E_J}$ have both started
with their break. The Fig.\ref{fig:10}(a) shows a
trajectory starting in the transient region close to NHIM $\mathcal{M}^3_{E_J}$ and going to NHIM $\mathcal{M}^1_{E_J}$.
And the Fig.\ref{fig:10}(b) shows a trajectory starting
on the transient part of NHIM $\mathcal{M}^1_{E_J}$ and ending on NHIM $\mathcal{M}^3_{E_J}$.

\subsection{The horizontal horseshoe}

The properties of homoclinic and heteroclinic connections can be illustrated best by the construction of horseshoes in some appropriate
Poincaré map. We have already seen in the previous subsection that horizontal heteroclinic trajectories help a lot to illustrate
heteroclinic connections between the various NHIMs. In addition, we can only plot Poincaré maps restricted to some invariant 2-dof
subsystem. Therefore, we construct now a Poincaré map for the horizontal subsystem and illustrate the creation of heteroclinic connections
between the NHIMs in this 2-dimensional map. For the construction of this Poincaré map, we use the intersection condition $y=0$ with
negative orientation. Remember that it is the same intersection condition which we have used before in the construction of Fig.\ref{fig:2}.
Therefore, we can relate important structures in the horizontal Poincaré map with particular periodic orbits contained in Fig.\ref{fig:2}
by just comparing the $x$ coordinates. The domain of the horizontal Poincaré map is the $x$ -- $p_x$ plane.

As already mentioned before, the only horizontal trajectories within the NHIMs associated with the collinear Lagrange points are the respective
horizontal Lyapunov orbits. And in the horizontal Poincaré map, they are represented by fixed points on the line $p_x=0$. Their
$x$ coordinate can be read off Fig.\ref{fig:2}. We use the Jacobi constant value $E_J=-0.043$ because this is the highest Jacobi constant where the
three horizontal Lyapunov orbits $lh_1$, $lh_2$, and $lh_3$ still exist, are still normally hyperbolic and still act as boundary curves of the
NHIMs, as explained in section 3. Therefore, it is clear that the corresponding fixed points in the horizontal Poincaré map are hyperbolic
fixed points. In the Poincaré plot, these three fundamental fixed points are marked by the black open circles and are called $lh_1$,
$lh_2$ and $lh_3$ respectively, exactly equal to the Lyapunov orbits which they represent.

\begin{figure}[h!]
\begin{center}
  \includegraphics[scale=1.0 ]{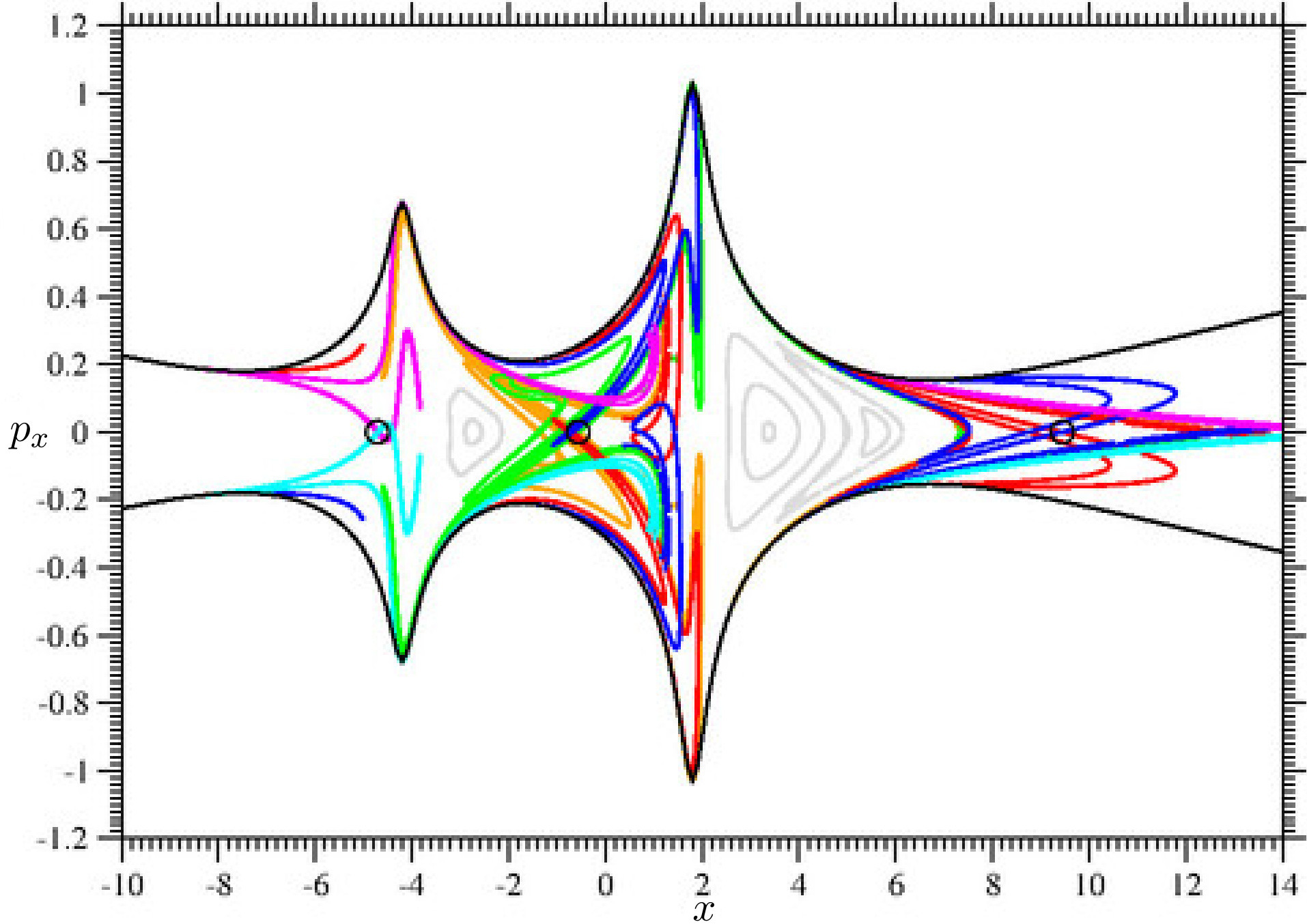}
\caption{Horseshoe at $E_J=-0.043$ in the invariant horizontal subsystem built up by the stable and
unstable manifolds of the horizontal Lyapunov orbits associated with the collinear Lagrange points.
In the Poincaré map,  $W^s(lh_1)$, $W^u(lh_1)$, $W^s(lh_2)$, $W^u(lh_2)$, $W^s(lh_3)$, and $W^u(lh_3)$ are plotted in the colours
orange, green, red, blue, magenta, and cyan, respectively. The fixed points representing the Lyapunov orbits are
marked by black open circles. The black curves are the boundary of the energetically accessible region at
$E_J=-0.043$. Some KAM islands around the minima of the effective potential  energy $V_{eff}$ are included as grey curves.}
\label{fig:11} 
\end{center}
\end{figure}

We calculate initial segments of the stable and unstable manifolds $W^s(lh_i)$ and $W^u(lh_i)$ of the three fundamental fixed
points and plot them into the $x$--$p_x$ plane, i.e. into the domain of the horizontal map. The result is Fig.\ref{fig:11}. From the trajectory
plots in the horizontal position space shown in the previous subsection, it is obvious that we have frequent violations of the
transversality of the intersection condition $y=0$. Therefore, the stable and unstable manifolds are torn into many disjoint fragments.
In the plot $W^s(lh_1)$, $W^u(lh_1)$, $W^s(lh_2)$, $W^u(lh_2)$, $W^s(lh_3)$ and $W^u(lh_3)$ are plotted in the colours orange, green, red,
blue, magenta and cyan, respectively. The black curves are the boundary of the energetically accessible region of the $x$--$p_x$
plane at $E_J=-0.043$.

Also included in grey colour are some KAM islands around tangentially stable periodic orbits. With the help of Fig.\ref{fig:2}, the
stable elliptic fixed points in the centres of these islands located at $x \approx -2.7$, $x \approx 3.4$ and $x \approx 5.8$
respectively can be identified with three horizontal periodic orbits growing out of the potential minima. 

In the figure, we make an interesting observation. Between the region around the Lagrange points $L_1$ and $L_2$, there exists 
a large KAM island which reaches almost to the energetic boundary of the domain of the map. This makes heteroclinic
connections to the NHIM $\mathcal{M}^2_{E_J}$ relatively difficult. In comparison, the KAM island between the regions
around the Lagrange points $L_1$ and $L_3$ is a lot smaller and leaves more room for heteroclinic connections to
NHIM $\mathcal{M}^3_{E_J}$. This observation makes it understandable that NHIM $\mathcal{M}^1_{E_J}$ has more
coordination of its bifurcation scenario with NHIM $\mathcal{M}^3_{E_J}$ than with NHIM $\mathcal{M}^2_{E_J}$.
Of course, for very high iterations of the initial segments of the stable and unstable manifolds, the ones from
any one of the NHIMs accumulate against the ones from all other stable or unstable manifolds, respectively,
of all invariant subsets of the dynamics. However, the corresponding heteroclinic trajectories might need a
very long time to come from the direct neighbourhood of the initial invariant subset to the neighbourhood
of the final invariant subset. On their way, these trajectories perform many complicated intermediate loops.

\subsection{Some remarks on transients and heteroclinic connections}

After having seen numerical evidence for transient effects and their influence on the general dynamics, let us finish this section
with some further remarks on the break of NHIMs and the related appearance of transient behaviour in the
region of the former NHIM. As long as the NHIM is a smooth 3-dimensional invariant surface of the flow in the 5-dimensional Jacobi constant
manifold (or a corresponding 2-dimensional surface in the 4-dimensional domain of the map), its stable manifold is a
smooth 4-dimensional surface in the Jacobi constant manifold (or a corresponding 3-dimensional surface in the map). It must have codimension 1 in
the Jacobi constant manifold and also in the domain of the map. Then a 1-dimensional curve $C$ in an appropriate region of the Jacobi constant manifold
close to the NHIM intersects the local branch of the stable manifold transversally. The curve $C$ can be chosen to lie in the domain
of the map, and then it also serves in the domain of the map for the same purpose. This is the basic idea behind the search strategy
for the NHIM as explained in all details in \cite{gj}.

Now imagine that, when we change the perturbation parameter, the NHIM loses its normal hyperbolicity in some region and breaks; only a lower-dimensional fractal collection of remnants of the NHIM remains. These remnants certainly contain unstable periodic orbits. The stable manifolds of these remnants have a lower dimension, i.e. higher
codimension than the stable manifold of the former complete NHIM. Then, in general, the curve $C$ will no
longer intersect exactly the stable manifolds of the remnants. Accordingly, the search strategy for the NHIM mentioned above will
no longer find these lower-dimensional remnants. However, if the curve $C$ is chosen in an appropriate region, it still contains
points leading to a high time delay in the neighbourhood of the remnants. This is the explanation of the transient effects.
To find the stable manifolds of such lower-dimensional remnants of NHIMs again, it would be necessary to scan initial
conditions not only along a 1-dimensional curve but on a higher-dimensional surface, where the dimension of this surface of initial conditions should be larger than or at least equal to the codimension of the stable manifold we want to catch. 

\newpage

\section{Visualisation of the phase space using the technique of delay time}
To complement our study of the phase space of the system as the perturbation parameter grows 
we are going to use a phase space structure indicator based on the delay time to visualise the invariant objects in the phase space.
The phase space indicator functions are a useful tool to analyse and get insights into the dynamics
of systems with many degrees of freedom. A review of the topic and several examples can be found
in the references \cite{sgl,wg,kag}. Also, recent studies of the phase space using Lagrangian descriptors
for a similar galactic model in 2-dof are presented in \cite{ts}. The delay
time is one of the best indicators to study the phase space of open systems due to its simple interpretation
and fast convergence; more information about it can be found in \cite{gj,gjman}.

Let us describe the construction of this phase space structure indicator function. First, 
we select a finite region $U$ of the 3-dimensional position space, which contains the position space projections of the important
unstable invariant sets which we want to investigate. In our case, we think of the three important NHIMs.
Let $X$ be any point in the phase space whose projection into the position space lies within $U$. 
Now we calculate the trajectory belonging to the initial point $X$
into the future direction of time and also into the past, in both time directions until the trajectory
leaves the region $U$ for the first time. Let us say into the future direction, it happens for the positive
time $t_+$, and for the past direction, it happens for the negative time $t_-$. Accordingly, the trajectory
stays within the region $U$ for the time $\Delta t = t_+ - t_-$. Then we define the time delay $t_d$ in $U$ as
$t_d = \Delta t - t_{fm}$, where $t_{fm}$ represents uninteresting parts of the
time in the asymptotic region where the particle behaves like a free particle.
Of course, for practical calculations, we introduce a time
limit $t_l$ and stop the calculation of the trajectory, if it does not escape from $U$ for a calculation time
$t_+$ smaller than $t_l$ in future direction or a calculation time $t_-$ larger than $-t_l$ in past direction.

\begin{figure}[h!]
    \centering
    \begin{tabular}{c c}   
     \subf{\includegraphics[scale=0.375]{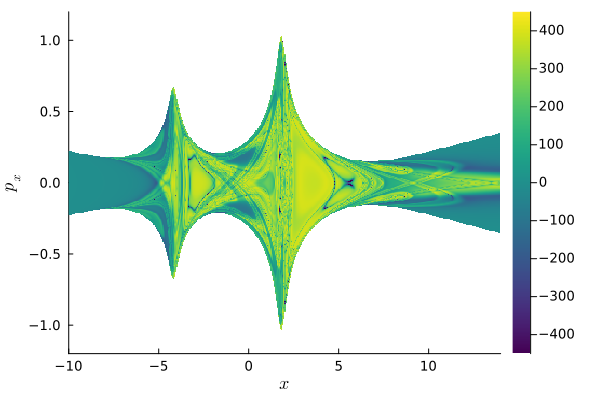}}
     {a) $E_J=-0.043$}
      &
     \subf{\includegraphics[scale=0.375]{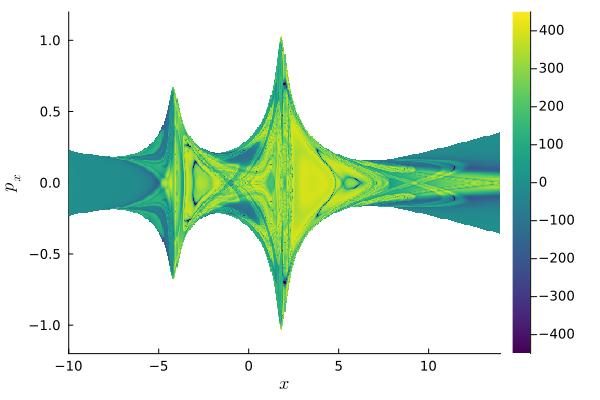}}
     {b) $E_J=-0.042$}
     \\
     \subf{\includegraphics[scale=0.375]{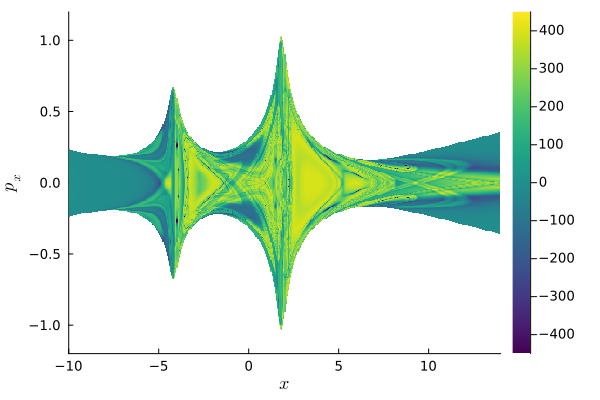}}
     {c) $E_J=-0.041$}
     &
     \subf{\includegraphics[scale=0.375]{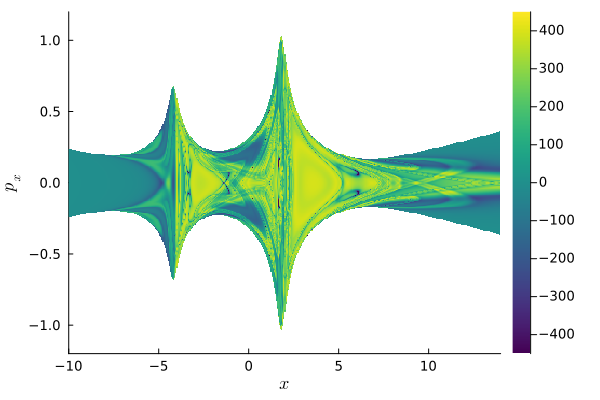}}
     {d) $E_J=-0.04$}
     \\
     \subf{\includegraphics[scale=0.375]{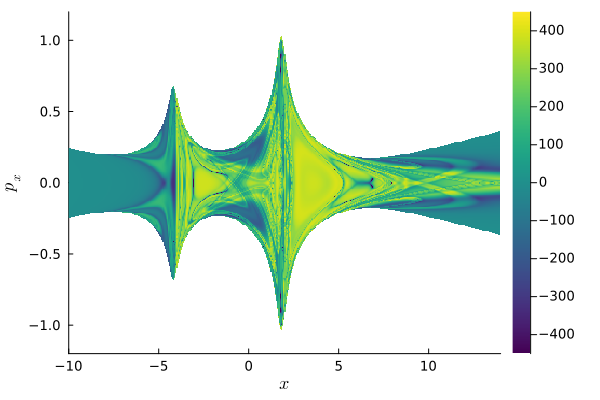}}
     {e) $E_J=-0.039$}
     &
     \subf{\includegraphics[scale=0.375]{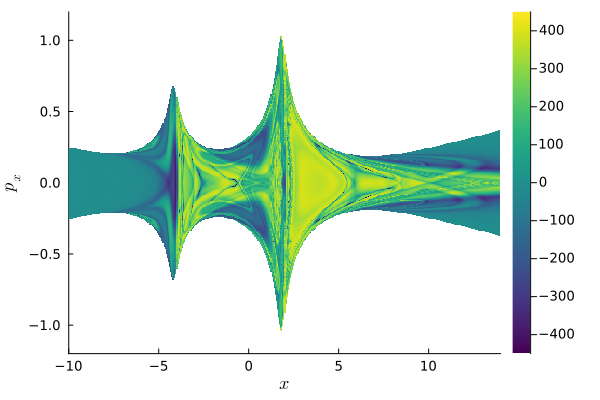}}
     {f) $E_J=-0.037$}
    \end{tabular}
    \caption{ Delay time indicator function in colour scale with initial conditions in the canonical plane $x$--$p_x$, $y=0$, $z=0$, and $p_z=0$ for different values of $E_J$.}
    \label{fig:12}
\end{figure}

To construct plots of the delay time indicator for our present system, we first chose as initial conditions
the canonical plane $x$--$p_x$ with $y=0$, $z=0$, and $p_z=0$, while the value of $p_y$ is determined
by the fixed value of the Jacobi constant $E_J$, where the negative branch of $p_y$ is chosen in order
to have the same orientation in $y$ direction as in Fig.\ref{fig:2} and Fig.\ref{fig:11}.
We select a fine grid of initial points on the $x$--$p_x$ plane and calculate
$t_d$ for each point. The result is presented in the Fig.\ref{fig:12} for 6 different values of $E_J$ from
the most interesting interval of $E_J$. The white regions on the plots correspond to energetically forbidden
initial conditions for the given value of $E_J$. Yellow points indicate large values of the
delay time $t_d$. These are points in stable KAM islands and also points lying on stable or unstable manifolds
of unstable invariant sets, in particular, the ones of the important NHIMs. Blue colour indicates points
which go to the asymptotic region rapidly in the future
and also in the past. 

As the value of $E_J$ increases, the structure of the KAM tori and of the homoclinic
and heteroclinic tangles changes but the NHIMs $\mathcal{M}^1_{E_J}$, $\mathcal{M}^2_{E_J}$, and $\mathcal{M}^3_{E_J}$ persist. Let us 
emphasise that the initial conditions for the plots on the panel
Fig.\ref{fig:12}(a) and Fig.\ref{fig:11} are the same. Therefore, we can appreciate the match between
the stable and unstable manifolds of the NHIMs and the stable KAM islands on both types of plots, i.e. on
Poincaré maps and on delay time indicator functions.

Comparing the panels in Fig.\ref{fig:12} as the value of $E_J$ increases, we notice that the value of the delay time $t_d$ decrees
on the region around the point $(x,p_x)=(-5,0)$, close to the NHIMs $\mathcal{M}^3_{E_J}$. This fact is evidence that the NHIMs 
$\mathcal{M}^3_{E_J}$ is losing hyperbolicity and the transient chaotic region close to it is growing in size as the value of $E_J$ increases.
This behaviour is expected when the value of $E_J$ increases far from the energy of the Lagrange point $L_3$, as we saw in previous sections. Similar behaviour of the delay time function indicator has seen in the references \cite{gjman,gj}. 

To continue with the analysis of the phase space of the system, we also consider initial conditions on the canonical
 plane $x$--$p_x$, $y=0$, $p_z=0$ and again a fixed value of the Jacobi constant $E_J=-0.043$, but now for different
initial values of $z$, namely
$z=1,2,3,4$. For $z \ne 0$, the plane of initial conditions does not belong to a lower-dimensional invariant subsystem.
Therefore, in general, trajectories do not return to this plane of initial conditions; the particle can move in all
directions in the energy shell of the phase space.

\begin{figure}[h!]
    \centering
    \begin{tabular}{c c}
     \subf{\includegraphics[scale=0.375]{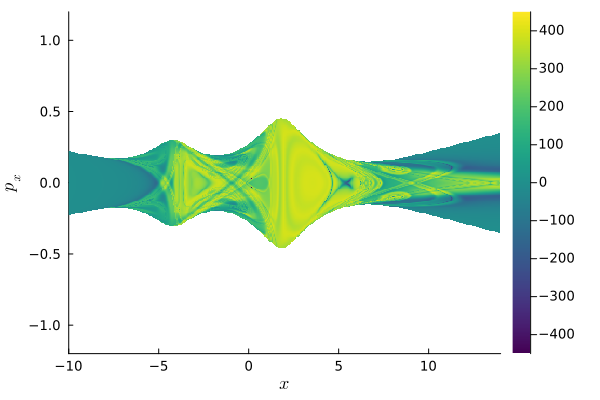}}
     {(a) $z=1$}
     &
     \subf{\includegraphics[scale=0.375]{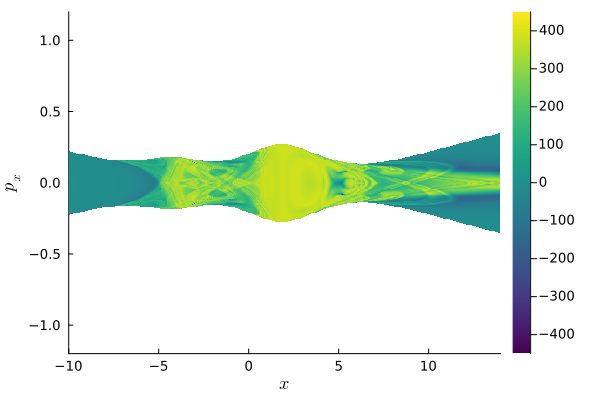}}
     {(b) $z=2$}
     \\
     \subf{\includegraphics[scale=0.375]{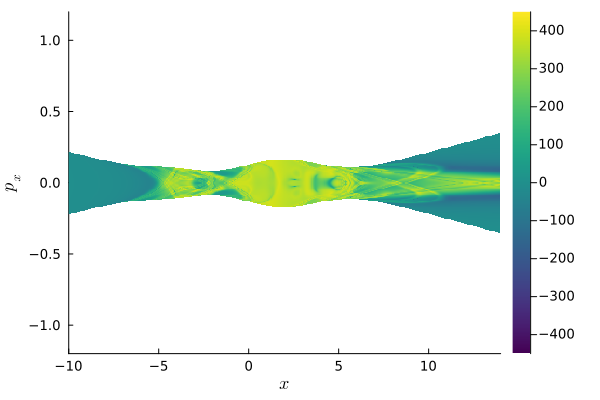}}
     {(c) $z=3$}
     &   
     \subf{\includegraphics[scale=0.375]{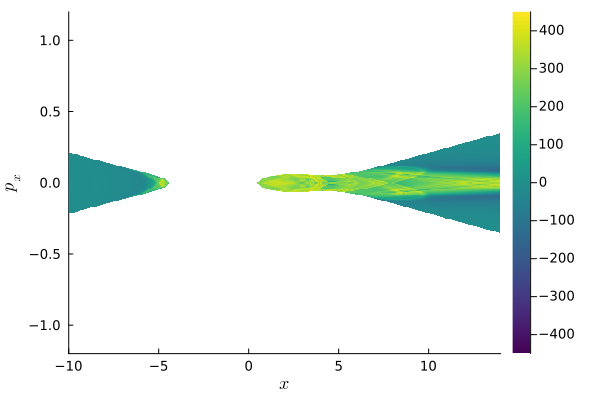}}
     {(d) $z=4$}

    \end{tabular}
    \caption{ Delay time indicator function for four different values of $z$ and for $E_J=-0.043$ }
    \label{fig:13}.
\end{figure}

The Fig.\ref{fig:13} displays the tangles of the NHIMs and the KAM tori for these sets of initial conditions.
The structure of the phase space looks qualitatively similar to the case with $z=0$ in Fig.\ref{fig:12}(a), we can
recognise the intersections of the stable and unstable manifolds of the three NHIMs $W^{s/u}(\mathcal{M}^1_{E_J})$,
$W^{s/u}(\mathcal{M}^2_{E_J})$, and $W^{s/u}(\mathcal{M}^3_{E_J})$ and the KAM tori. In addition, we can see a qualitative
change of the energetically allowed region for $z=4$; here, the set of energetically accessible initial conditions
has two disjoint components.

From Fig.\ref{fig:13}, we notice that the KAM tori associated with the point $P_2$ intersects the planes with different 
value of $z$, forming a barrier in the phase space. This KAM tori gives little room for the trajectories to 
travel between the neighbourhood around the NHIM $\mathcal{M}^1_{E_J}$ and the other two NHIMs $\mathcal{M}^1_{E_J}$ and $\mathcal{M}^1_{E_J}$,
as we mention before in previous sections.

\newpage

\section{Conclusions and final remarks} 

The system analysed in the present article is an instructive example of Hamiltonian dynamics with 3 not symmetry-related NHIMs, and
for the effects of coordination of the bifurcation scenario of the NHIMs. A new feature in the present article is the inclusion of
transient behaviour close to NHIMs generated by their break into the formation of the global scenario. We saw indications that also the transient chaotic trajectories around the NHIMs give important contributions to the global dynamics.

In the study of the bifurcation scenario of the NHIMs and the comparison with the corresponding left-right symmetric system
investigated in \cite{zj1}, we made the following interesting observation: Also, when the symmetry between the two outer NHIMs is broken by the transition to different masses of the two dwarf galaxies, there remains a partial qualitative symmetry in
the bifurcation scenarios of the two outer NHIMs. 
This partial symmetry includes the sequence of the important bifurcations and their
order for the Lyapunov orbits contained in the individual NHIMs. But it does not include the coordination of the creation of chaos.
Here, in this example, only NHIM $\mathcal{M}^1_{E_J}$ and NHIM $\mathcal{M}^3_{E_J}$ make coordinated bifurcations into the
direction of chaos; NHIM $\mathcal{M}^2_{E_J}$ does not participate in this coordination. This happens even
though there are heteroclinic trajectories between any pair of NHIMs. The horizontal Poincaré
map presented in subsection 4.2 gives at least some indication of the relatively weak heteroclinic coupling of NHIM
$\mathcal{M}^2_{E_J}$ to the other two fundamental NHIMs of the system.

It seems that NHIM $\mathcal{M}^1_{E_J}$ and NHIM $\mathcal{M}^3_{E_J}$ become susceptible to tangential instabilities
at a similar Jacobi constant interval, while NHIM $\mathcal{M}^2_{E_J}$ is not yet ready for the development of large-scale
chaos. Of course, in the symmetric case studied in \cite{zj1}, NHIM $\mathcal{M}^2_{E_J}$ has exactly the same bifurcation
scenario as NHIM $\mathcal{M}^3_{E_J}$, and the two NHIMs develop large-scale chaos in the same Jacobi constant interval.
The horizontal horseshoe plot of Fig.\ref{fig:11} suggests that the large-scale KAM island associated with the potential minimum
$P_2$ prevents a rapid and efficient heteroclinic connection between NHIM $\mathcal{M}^1_{E_J}$ and NHIM
$\mathcal{M}^2_{E_J}$ along horizontal trajectories. In our collection of heteroclinic orbits, we do not have a single
one which starts or ends on NHIM $\mathcal{M}^2_{E_J}$, runs through the potential minimum at $P_2$ and
does not go to $z$ values far away from $z=0$. As we have seen in subsection 4.1, long and complicated trajectories can do it.
As Figs.\ref{fig:8}(e) and \ref{fig:9}(a) show, there are also trajectories going to the outside and circling around $L_4$ or $L_5$
can do it. But these long trajectories seem to be little efficient in the creation of coordination effects in the
bifurcation scenarios between NHIMs.
Of course, also trajectories using the vertical degree of freedom can establish heteroclinic connections, remember the
trajectory from Fig.\ref{fig:9}(d). However, we have seen here (and also in other model systems) that in systems with a
$z$ reflection symmetry and accordingly an invariant horizontal subsystem, the energy exchange between the vertical degree of
freedom and the horizontal degrees of freedom is rather slow. This prevents an efficient heteroclinic coupling between the NHIMs via
trajectories having a large part of their Jacobi constant in the vertical motion.

One important aspect of the bifurcation scenario, which we can not yet explain precisely, but which certainly
deserves a further, more fundamental investigation, is the following consideration: For stable and unstable
manifolds of NHIMs, there holds a foliation theorem which loosely speaking says that these manifolds are the
Cartesian product of a line with the internal NHIM structure. Thereby, these manifolds transport the internal
NHIM structure along with them to all regions of the phase space which are heavily influenced by the homoclinic/heteroclinic
tangle of the NHIMs. As a consequence, the whole homoclinic/heteroclinic tangle is at least qualitatively
similar to a Cartesian product of a lower-dimensional homoclinic/heteroclinic tangle (in our case, the horizontal horseshoe presented
in subsection 4.2) and the internal NHIM structures. This property has been noticed before and investigated in more detail
in \cite{zj3}. Also, the system investigated in this reference is 3-dof and has, in a natural form, an invariant horizontal 2-dof
subsystem. Such a tendency of higher-dimensional homoclinic/heteroclinic tangles to show an approximate product structure
has also been observed by another group in \cite{dre1, dre2, dre3}. This tendency seems to be common and needs a more
profound explanation. 

The appearance of the tangential transient effects shows that, also after the beginning of the break NHIMs, its remnants, the transient region formed close to the NHIM, have an effect on their environment in the phase space and on the formation of trajectories with similar behaviour to the homoclinic/heteroclinic tangle that existed before the break of the constituent KAM tori of the NHIMs. The fragments and transient parts continue without heavy qualitative differences. 

The NHIMs associated with the Lagrange points of the effective potential direct the transport of test particles through these effective potential saddles. 
In the interpretation of our model as describing two dwarf galaxies in interaction, this model describes the exchange of
stars between the two galaxies (motion through the potential saddle associated with the central Lagrange point $L_1$) or the loss of stars from the binary galaxy
system (motion through the potential saddles associated with the outer Lagrange points $L_2$ and $L_3$). The projection of the unstable manifolds of the NHIMs into
the position space then gives the position of tidal arms and tidal bridges going out from the dwarf galaxies. Of course, our
model system could also be interpreted as a simple model for two nearby globular clusters in interaction. Here, it is interesting
to mention that in the astronomical literature, there is a claim of observational indications for a tidal bridge between the
interacting globular clusters NGC5024 and NGC5053 \cite{chun}. 

Finally, let us remark that the system studied is a variant of the Lagrange three-body problem. 
Therefore, it has similar properties to those of other variants of the restricted three-body system. 
Some main qualitative properties of the phase space found in the present work are rather robust against perturbations of the system. 
This can be seen first from the similarity with the symmetric case studied in detail in \cite{zj1}.
And second, the persistence theorems for NHIMs and their invariant stable and unstable manifolds guarantee that 
their chaotic invariant set will be similar for a general small perturbation of the system. Small perturbations of the system might 
be small modifications of the gravitational potential, for  example, to come to the anisotropic Kepler problem, and they could also 
include relativistic corrections. Also, we expect that our system has similarities with the galactic model recently studied in \cite{harsoula}.

\section*{Acknowledgment}

We thank DGAPA-UNAM for financial support under grant number IG-101122 and CONACyT/SECIHTI for financial support under grant number 425854.

\newpage

\section{Appendix: Algorithm to calculate the NHIM using stabilisation of trajectories by the delay time}

In the present work, the algorithm we use to calculate an approximation to the NHIM and its internal dynamics
is based on the behaviour of the trajectories in the neighbourhood of the NHIM and its
geometry.  This method is a direct multidimensional generalisation of the classical algorithm of control of chaos for
the stabilisation of unstable periodic orbits using its stable manifold and the Poincaré map, which ensures
convergence to the periodic orbit. In the general
multidimensional case, we construct segments of trajectories
which belong to the stable manifold of the NHIM and are already
very close to the NHIM itself. They are excellent approximations
to trajectories running within the NHIM. If we construct a
sufficiently large number of such trajectory segments
that converge to the NHIM, we can obtain an approximation to the NHIM and its internal dynamics.

To explain the basic steps of the algorithm, let us consider a Hamiltonian system defined by the Hamiltonian function 
$H=T(p)+V(q)$ with $n$ degrees of freedom. Suppose that the potential energy $V(q)$ has one index-one saddle point 
$q_s= (q_{s1},q_{s2},..., q_{sn})$. In the configuration space, this equilibrium point has one unstable direction
$q_1$ and $n-1$ stable directions $q_2,...,q_n$ for the dynamics. 
In the phase space, this saddle point corresponds to an unstable fixed point with the energy $E_s=V(q_s)$. 
Suppose that for an energy $E>E_s$ there is a $2n-3$ dimensional NHIM $\mathcal{M}_{E}$,
then exist its $2n-2$ dimensional stable and unstable manifolds $W^{s/u}(\mathcal{M}_{E})$ \cite{wiggins1,wiggins2}.
Let us remark that the stable and unstable manifolds $W^{s/u}(\mathcal{M}_{E})$ divide the $2n-1$ dimensional constant energy manifold; this fact is very important for our algorithm because it helps us to find an approximation to the NHIMs using simple geometrical ideas.

\begin{figure}[h!]
    \centering
    \begin{tabular}{c c}
     \subf{\includegraphics[scale=0.375]{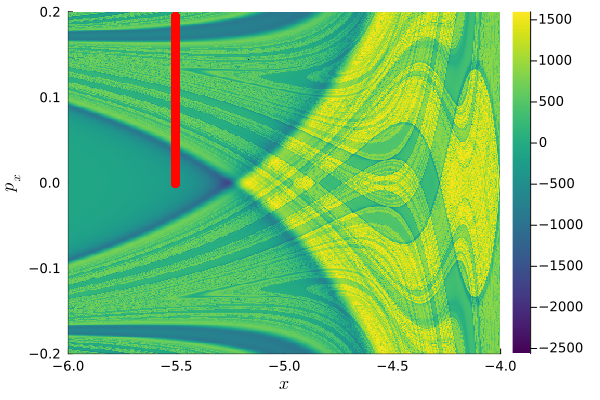}}
     {(a) $t_d$ evaluated on the canonical plane $x$--$p_x$.}
     &
     \subf{\includegraphics[scale=0.375]{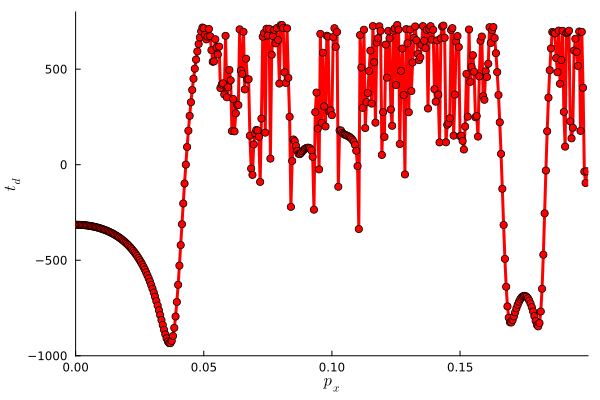}}
     {(b) $t_d$ evaluated on the red line segment.  }
     \\

    \end{tabular}
    \caption{ Delay time indicator function and the algorithm to find NHIMs. Part (a) shows the delay time function $t_d(x,p_x)$ plotted in colour scale and evaluated on a region that intersect the NHIM $\mathcal{M}_{E+\Delta E}$. The intersection of the NHIM $\mathcal{M}_{E+\Delta E}$ with the initial conditions is the corner of the big smooth curved triangle around $(x,p_x)=(-5.25,0)$. The upper side of this triangle is the intersection of a segment of the stable manifold $W^{s}(M_{E+\Delta E})$ with the set of initial conditions. The singularities of $t_d(x,p_x)$ are indication of $W^{s/u}(M_{E+\Delta E})$. Part (b) $t_d(x=-5.5,p_x)$ is the value of the time indicator function on the red line on part (a), the first singularity from the left to the right side is a good approximation of the point  $X^* \in W^{s/u}(M_{E+\Delta E})$.  }
    \label{fig:14}.
\end{figure}

Now, let us suppose that for an energy $E+\Delta E$ there is a NHIM $\mathcal{M}_{E+\Delta}$ with the same 
dimensions as $\mathcal{M}_{E}$, then exist the stable and unstable manifolds $W^{s/u}(\mathcal{M}_{E+\Delta E})$ 
with the same dimensions as $W^{s/u}(\mathcal{M}_{E})$. The basic steps to search points in the NHIM $\mathcal{M}_{E+\Delta E}$ are the following:

\begin{enumerate}
    \item[I.]  Take a point $X=(q^\bigstar_1,q^\bigstar_2,...q^\bigstar_n,p^\bigstar_1,p^\bigstar_2...,p^\bigstar_n) \in \mathcal{M}_{E} \bigcap \Sigma$,
           where $\Sigma$ is the domain of Poincaré map.
    \item[II.] Take the projection of the point $X$
          on the hyperplane $p_1 =0$ defined by the canonical momentum conjugate to the unstable direction  $q_1$, lets denote this point as $X_{0}=(q^\bigstar_1,q^\bigstar_2,...q^\bigstar_n, 0,p^\bigstar_2...,p^\bigstar_n)$.
    \item[III.] Take a point $X_{p^\bigstar_1+\Delta p_1} =(q^\bigstar_1,q^\bigstar_2,...q^\bigstar_n,p^\bigstar_1+\Delta p_1,p^\bigstar_2...,p^\bigstar_n)$, where the increment $\Delta p_1$ is enough to intersect the $W^{s}(\mathcal{M}_{E+\Delta E})$.
    \item[IV.] Calculate $l(X_{0}, X_{p^\bigstar_1+\Delta p_1})$ the line segment that starts at  $X_{0}$  and finishes at $X_{p^\bigstar_1+\Delta p_1}$, see Fig.\ref{fig:14}(a). The points in this line segment are given by $X_{p_1}=(q^\bigstar_1,q^\bigstar_2,...q^\bigstar_n,p_1,p^\bigstar_2...,p^\bigstar_n)$, where $p_1 \in[0,p^\bigstar_1+\Delta p_1]$.

    \item[V] Calculate the delay time function  $t_d(l(X_{0}, X_{p^\bigstar_1+\Delta p_1}))$, see Fig.\ref{fig:14}(b).
    \item[VI.] Find the first singularity of the delay time function $X^*\in l(X_{0}, X_{p^\bigstar_1+\Delta p_1})$ such that 
    \[  \lim_{ X \to {X^*}^+ }   t_d(X) =  \lim_{ X \to {X^*}^+ } t - t_{fm}  = \infty .\]

where $t$ is the integration time and $t_{fm}$ is the time that free particles take from the origin
to the position $q(t)$. 

To approximate $X^*$ numerically, we can use the bisection method with the conditions $ \frac{d t_d(X_{p_1}) }{dp_1} < 0 $ and $t_d^{max} > |t_d(X_{p_1})|$ to approximate the singularity $X^*$ from the left and avoid the region with an infinite number of singularities, see  Fig.\ref{fig:14}(b). The point $X^*$ is a good approximation to a point in $W^{s}(\mathcal{M}_{E+\Delta E})$ in the neighbourhood of $\mathcal{M}_{E+\Delta E}$.
    \item [VII.] Calculate the Poincaré map $P(X^*)$.
    \item [VIII.] Repeat the previous steps (I--VI) using the point $P(X^*)$ to obtain an approximation of the next iteration of the Poincaré map.
    \item [IX.] If we can obtain a very large number of iterations of the point $X^*$ using this procedure of stabilisation, we consider
          that all iterations of $X^*$ could be part of the $\mathcal{M}_{E+\Delta E}$.
\end{enumerate}

In case the algorithm fails to find the singularity for the delay function $t_d$, it is likely that $W^{s}(\mathcal{M}_{E+\Delta E})$
and $\mathcal{M}_{E+\Delta E}$ do not exist in a neighbourhood of the initial point considered in the algorithm.

\end{document}